\documentclass[12pt,preprint]{aastex}

\newcommand{\Teff}{\mbox{$T_{\rm eff}$}}

\newcommand{\Msun}{\mbox{$\cal M_{\odot}$}}

\newcommand{\MV}{\mbox{$M_V$}}

\newcommand{\as}{\mbox{$^{\prime\prime}$}}

\newcommand{\ebv}{\mbox{$E(B-V)$}}

\newcommand{\mv}{\mbox{$m_{V}$}}
\newcommand{\magf}{\mbox{$m_{170}$}}
\newcommand{\magn}{\mbox{$m_{255}$}}
\newcommand{\magu}{\mbox{$m_{336}$}}
\newcommand{\magb}{\mbox{$m_{439}$}}
\newcommand{\magv}{\mbox{$m_{555}$}}
\newcommand{\sqamin}{\mbox{$arcmin^{2}$}}
\shortauthors{Bianchi \& Efremova}
\shorttitle{HST Imaging of NGC 6822}   
\slugcomment{Version of  \today}

\begin{document}

\title{The recent star formation in NGC 6822 from HST imaging
\footnote{based on observations made with the NASA/ESA {\it Hubble Space telescope},
obtained at the Space Telescope Science Institute, which is operated by the Association
of Universities for Research in Astronomy, Inc., under NASA contract NAS 5-26555.} }
\author{Luciana Bianchi\altaffilmark{1} and Boryana V. Efremova\altaffilmark{1,2} 
\altaffiltext{1}{Center for Astrophysical Sciences, 
Department of Phys.\& Astron., The Johns Hopkins
University,3400 N.Charles St., Baltimore, MD21218
(bianchi@pha.jhu.edu, boryana@pha.jhu.edu)}
\altaffiltext{2}{Department of Astronomy, University of Sofia, 5 James Bourcheir St., Sofia 1126, Bulgaria}
}

\begin{abstract}
We present HST WFPC2 and STIS  imaging of 
the low metallicity galaxy NGC\,6822, 
performed as  part of a study of the young  stellar populations in the 
galaxies of the Local Group.
Eleven WFPC2 pointings, with some overlap, cover two regions, 
extending over  19 \sqamin ~and 13 \sqamin ~ respectively,
off the galaxy center.
The filters used are F170W, F255W, F336W, F439W, and F555W. 
One 25$\times$25$''$ field observed with STIS' FUV- and NUV- MAMA, 
includes Hodge's OB8 association and
the HII region Hubble V, contained in  
Field 1 of Bianchi et al.: this
previous study provides additional WFPC2 four-band photometry.  
 
We derive the physical parameters of the stars in the fields and the
 extinction  by comparing 
the photometry to grids of  model magnitudes.
The environments studied in this work include one of the most luminous
(in H$\alpha$) HII regions in the Local Group (Hubble V) with a compact star cluster, 
a typical OB association (OB15), the sparse
field population and the outskirts of NGC6822. 
In the WFPC2 fields, most of the 
hot massive stars are found  in the Hodge OB15 association,
at about 5$'$ [0.7kpc] East of the galaxy center,  
extending  $\approx 90''$ [$\approx 200 pc$] in our imaging.
The  color-magnitude
diagram indicates a young age, $\leq$ 10Myrs, for this association, where 
we measure 70 stars 
hotter than $\sim$ 16000~K (earlier than
mid-B spectral type ) according to their photometric colors.
In the compact HII region Hubble V, we measure 80
stars brighter than 
 m$_{NUV}$ $<$ $22^{m}.5$, most of them have high temperatures.
The density [per unit area] of hot massive stars in the core of the OB8 association 
is higher than in OB15 by a factor of 12,  while the total stellar
mass formed is similar ($\approx$ 4 or 7 10$^3$ \Msun, 
when extrapolated to a 
mass range of 1-100\Msun ~ or  0.1-100\Msun respectively).
In both OB15 and OB8  massive star candidates are found.
In the general field outside of the OB15 association, 
we find  few hot massive stars (0.7/arcmin$^2$) and 
several A-type supergiants.
No massive star candidates are found in the WFPC2 fields outside
the main galaxy body ($\approx$ 10.5-14.5$'$ [1.5-2.kpc] from the galaxy center),
where the population is dominated by foreground stars,
 at least down to $V\sim 22$.   At fainter magnitudes, we measure
in these outer fields 
a significantly larger number of stars than the model for MW foreground
objects would predict. 
The average extinction is found to vary among the three environments
studied: \ebv = 0.22 in the outer
regions, \ebv = 0.27 in the fields East of the galaxy main bar, and 
\ebv = 0.40 in the HII region  Hubble V. 

A quantitative 
discussion of the applicability of the reddening-free-index  method
for photometric determination of stellar parameters
is provided for the filters used in this work, based on our grids of stellar models.  
\end{abstract}




\section{Introduction}

The massive star populations of nearby galaxies are a subject 
of great interest, due to the impact that massive stars have on the galaxies' 
properties and evolution.  Massive stars, in 
their short lifetime, drive the chemical and dynamical evolution of 
the parent galaxies and trace the star--formation activity.
The most luminous, most massive stars provide information on the Initial Mass 
Function (IMF) and its properties. The study of the IMF in different 
environments 
gives us a way to determine its dependence on parameters such as metallicity 
and star--formation rate.
 The galaxies of the Local Group provide an excellent opportunity to 
investigate the modalities of star formation through the resolved studies of their 
stellar populations, because they are close enough so that individual stars 
can be accurately measured, and all the stars are at approximately the 
same, known  distance thus their absolute luminosity can be derived.
With its high resolving power, and wide range of filters including  UV
wavelengths, the {\it Hubble Space Telescope} (HST) is the most suitable 
instrument for studying young 
stellar populations in the Local Group galaxies.
In a global context, studies of resolved
local starbursts provide a  key to understand distant star-forming 
galaxies, for which only integrated properties can be measured.
Resolved studies of the stellar content, extinction and global parameters of 
nearby starburst regions in different
environments provide a calibration to  interpret integrated studies
 of unresolved starbursts in distant galaxies.

NGC\,6822 is an irregular dwarf galaxy, a member of the Local Group.
The angular size of the galaxy at optical wavelengths is $6'\times11'$ 
and its distance is estimated to be 500 kpc (McGonegal at al. 1983).

The old and intermediate stellar populations of NGC\,6822
have been  studied by Gallart et al. 
(1996) on a central 11.2$'$ $\times$ 10.4$'$ area, and by 
Wyder (2003) in four HST-WFPC2 fields, from V and I band photometry.  
Both studies infer from the CMD that the star formation began
12-15Gyr ago or more recently, depending on the initial metallicity
assumed, and proceeded more or less uniformly until about 600Myr ago.
The young stellar population of NGC\,6822, and its recent star formation
history,  has also been the subject of previous 
photometric studies. The first galaxy--wide photographic UBV photometry of
 NGC\,6822 was published by Kayser (1967), followed by a photographic and 
photoelectric study by Hodge (1977), who catalogued sixteen OB associations. 
Massey et al. (1995) ground-based photometric study 
found that O-type stars are present in several of the OB associations,
but that  NGC\,6822 is 
relatively poor in very massive stars compared to M31, M33 and the Magellanic 
Clouds; 
 the reddening varies from \ebv=0.26 on the periphery to 0.45 near 
the central regions, with an average ratio $E(U-B)/E(B-V)$ of $0.74 \pm 0.07$.
Bianchi at al. (2001a) presented ground--based UBV CCD photometry covering the 
whole body of the galaxy and  HST Wide Field Planetary Camera 2 
(WFPC2)  four band photometry in two regions 
containing very rich and crowded OB associations. They found populations 
younger than 10 Myr in four of the OB associations included in their study, 
and  in particular  
 very young stellar populations (few Myr) in the star--forming 
HII regions Hubble V and Hubble X (Hubble 1925). 
The outer parts of the galaxy are much less studied. 
No young stellar population was found 
in the two outer regions of NGC\,6822 analyzed by  
Hutchings, Cavanagh, \& Bianchi (1999).

The search for young stars in different galaxy environments presented in 
this paper covers larger regions than previous HST studies, with 
deeper exposures, and has the advantages of the HST wide wavelength coverage
(extending to the UV), 
and higher spatial resolution, compared to the ground-based UBV data.
 The  UV bands (WFPC2 F170W and F255W, {\it Space Telescope
Imaging Spectrograph} (STIS) MAMA-FUV and
MAMA-NUV) in addition to the U, B, V bands, allow us to identify the most 
massive, luminous stars. In fact, due to their high effective temperatures,
the earliest spectral types cannot be differenciated from 
optical colors (see e.g. the discussion in Massey (1998a) and 
Bianchi \& Tolea, 2006, in preparation).

The observations and data reduction are described in section 2, the 
photometry measurements and the characteristics of the 
stellar populations are presented in section 3, 
   the results are discussed in section 4, and summarized in 
Section 5.

\section{Observations and Data Reduction. Photometry.}
\label{sdata}

Hubble Space Telescope (HST) imaging of different regions in  NGC\,6822 were 
obtained between 6/16/2000 and 5/24/2002 as  part of the program GO 8675 
(P.I. L. Bianchi) to study the massive star content of this galaxy, with the 
Wide Field and Planetary Camera 2 (WFPC2) and the 
Space Telescope Imaging Spectrograph
(STIS) imagers. The WFPC2 images were taken in parallel mode. 
We discuss the reduction and analysis procedures separately
for the two sets of data. The photometry and the derived parameters for
the sample stars are given in Table 3 (printed for the hottest stars subsample, 
in electronic form for the entire sample), and 4 (STIS data). 

\subsection{WFPC2 imaging}
\begin{figure}
\hskip -1.cm
\plotone{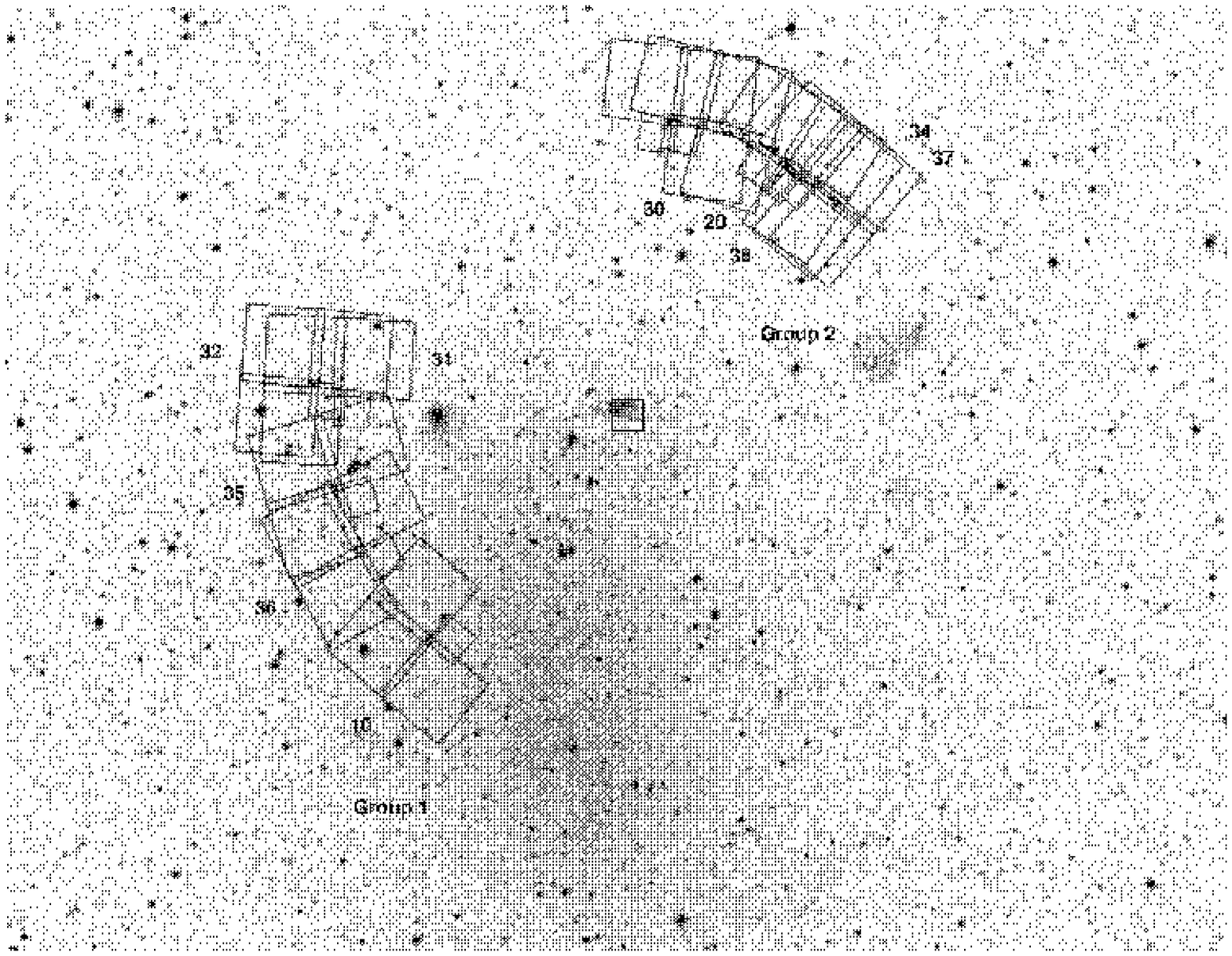} 
\caption{Ground-based V-band image ($\approx$ 21$'$ $\times$ 16 $'$ )
of NGC 6822 with outlined WFPC2 and STIS fields. 
North is up, East is to the left.  The small square between the two WFPC2 
groups indicates the STIS field. 
\label{fig:fld} }
\end{figure}

 A total of 110 images in 11 different fields were obtained  with WFPC2 
on-board HST. The instrument consists of four $800 \times 800$ CCD chips 
with angular resolution of $0\arcsec.046\, pix^{-1}$ [0.11pc at the distance of 500 kpc to NGC6822] for 
the Planetary 
Camera (PC) and $0\arcsec.099\, pix^{-1}$ [0.24pc] for the Wide Field chips 
(WF2, WF3 and WF4). In most of the fields imaging was taken with 
five WFPC2 broad -- band filters: F170W, F255W, F336W, F439W and F555W. 
Table~\ref{tbl:flds} lists the field positions, 
the filters used and the corresponding 
exposure times. The fields are partially overlapping, and are
 grouped in two areas, East and North-West the main body 
of the galaxy. The East group (Group 1), covers 
about  19 \sqamin [$\approx$ 402kpc$^2$] , and 
the N-W group (Group 2), covers about 13 \sqamin [$\approx$ 275 kpc$^2$]. 
The positions of the fields are shown in Figure~\ref{fig:fld}.

The images processed with the Post Observation Data Processing System (PODS)
 pipeline for bias removal and flat fielding were downloaded from the MAST 
archive.
In order to reject the cosmic rays, images taken with the same pointing 
and filter were combined using the IRAF task CRREJ. This task combines 
multiple exposures at the same pointing and rejects the high counts that 
occur in only one of the frames. It was possible to apply this procedure to 
all the fields, because we have at least two images per filter for each 
pointing.
The Source Extractor code (Bertin and Arnouts 1996) was then used to detect the 
star-like objects exceeding the local background by more than 
$3.5~\sigma$ in the F555W images. Then aperture photometry was performed, 
using the same positions in all the filters. Slight recentering 
(less than 2 pixels) from the F555W position was needed to optimize centering
of the objects from filter to filter.
We performed the photometry with  the PHOT task in the DAOPHOT/IRAF package,
using the source list and  positions from Source Extractor. 
The aperture chosen for both PC and WF chips was 3 pixels, and the sky was 
 measured in an annulus with inner and outer radii of 5 and 8 pixels 
respectively.
Aperture corrections to 0\arcsec.5 aperture were measured using isolated 
stars separately for the different chips and filters.
Aperture photometry was chosen over psf photometry 
to ensure a uniform procedure among our fields, and consistent
with previous works. Crowding effects are negligible in our fields.
The WFPC2 fields in particular contain fairly sparse
populations (the most crowded region is shown in Fig.6), and the only crowded region, Hubble~V, was
observed with STIS at UV wavelengths, where crowding is much less than at optical
wavelengths.   
  The flux calibration was performed following Holtzman et al. (1995a,b) 
with the refinements of Dolphin (2000), and included the following steps: 
(i) Distortion Correction:  geometric distortion affects the integration 
in the photometric aperture.
To account for this effect our images were multiplied by the correction 
image provided by Holtzman et al. (1995b). 
(ii) Correction for Contamination:  due to contaminants buildup on the cold 
CCD plate the UV throughput decreases. The photometric correction to be 
applied depends on the time elapsed since the last decontamination procedure. 
Decontamination correction coefficients were taken from McMaster \& Whitmore 
(2002). (iii) Charge Transfer Efficiency (CTE) correction: the formula of 
Dolphin (2000) was applied.

 The magnitudes were then transformed in the HST VEGAMAG photometric 
system by applying the zero points as follows: 
for F336W, F439W and F555W, we used the zero points 
 from Dolphin's (2002) web site, which are relative to a 0\arcsec.5 aperture.
The zero points for F170W and F255W were taken from the HST Data Handbook 
for WFPC2 with a $0^{m}.1$ correction for the transition from infinite 
to 0\arcsec.5 aperture. For our analysis we used the VEGAMAG WFPC2 
photometric system, but in Table 3 (the hottest stars)  the F336W, F439W and F555W 
magnitudes are also 
transformed to the Johnson's UBV system to allow comparisons with other
works.  The error vs magnitude plots of a 
typical field (FIELD 10) for all the filters used are shown in 
Figure~\ref{fig:err}.
Such diagrams are used to estimate the magnitude limits
of our sample after we impose error cuts for the analysis described
in Section \ref{s_analysis}.

The overlapping fields were processed separately and photometry was performed 
in each field independently. The measured magnitudes of stars included 
in several fields are consistent within the errors. The magnitude given 
in the final catalog is an error weighted average of all the measurements for the 
same star. 
The formula used is  
$\overline{m}=\frac{\sum_{i=1}^{n} \frac{m_i}{\Delta m_i^2}}{\sum_{i=1}^{n} \frac{1}{\Delta m_i^2}}$, 
where n is the number of overlapping fields in which the star is present. 
The corresponding error is calculated as: 
 $\overline{\Delta m}=(\sum_{i=1}^{n}\frac{1}{\Delta m_i^2})^{-1/2}$

The deepest observations are in the F555W band, where 18244 stars are 
detected (above 3.5$\sigma$); 12423 (70\%) of them have photometric accuracy better than 
$0^{m}.2$. The number of stars with errors smaller than $0^{m}.2$ in 
the F439W filter is 1757 and in F336W the number of stars 
with accuracy better than $0^{m}.2$ is only 535 (2.7\%   of the stars detected in F555W). 
Fewer stars are detectable in F170 and F255 with accuracy better than 
$0^{m}.2$. See Table~\ref{tbl:errs} for details on errors and magnitude 
limits in the different filters.

\subsection{STIS imaging}

Four overlapping STIS 
images were taken with the 25MAMA aperture using the 
 FUV-MAMA and NUV-MAMA photon counting detectors. 
All the images are centered at $RA\approx 19\,44\,52.2$ and 
$Dec\approx  -14\,43\,14.5$ and cover a $25''\times25''$ [60 $\times$ 60~pc]
field of view. 
The size of the detectors is $1024\times1024\,pixels$, 
and the resolution is $0''.024 pixel^{-1}$ [0.06~pc], providing the
 highest resolution view of the compact HII region Hubble~V to date. 
Details about the images are listed in Table~\ref{tbl:flds}.
The STIS data were downloaded after being calibrated through the STScI 
{\it calstis} pipeline. Aperture photometry (5 pixels) was performed using the 
PHOT task in  the DAOPHOT/IRAF package. 
The aperture corrections were measured from 20.8 pixel aperture 
(corresponding to $0''.5$) photometry of selected stars.
The measured magnitudes were converted to the HST VEGAMAG system.
There are 80 stars detected in both NUV and FUV.
The magnitude - error diagram is shown in Figure~\ref{fig:st_err},
and the magnitude limits are listed in Table ~\ref{tbl:errs}.
Our entire STIS field is included in WFPC2 Field 1 of Bianchi et al. (2001a).
This previous work provides photometry in four WFPC2 filters, F555W, 
F439W, F336W
and F255W, that will be combined with the UV STIS
photometry in the analysis described in the next section.
The STIS images contain the OB8 association (Hodge 1977) embedded in the 
HII region Hubble V. 
The STIS imaging provides a gain of over 2 magnitudes in NUV with respect
to the earlier WFPC2 data, and the first observations of this region
in the FUV. 
We have six band photometry (four bands from Bianchi et al. 2001a WFPC2 photometry 
and two from our STIS photometry) for 72 out of the 80 stars  
 detected in our STIS imaging.

\section{Analysis. The Stellar Populations}
\label{s_analysis}

In this section we analyze  the multi-band photometry of individual stars 
and derive their physical  parameters by comparison with model colors. Two methods are used,
and the results are compared. For the stars with good photometric 
measurements  in all the filters,
we fit all the observed colors with model colors, to which various 
reddening amounts are applied, and derive simultaneously stellar
\Teff ~ and interstellar extinction. For the stars not detected in 
the UV filters, or with
large uncertainties in the UV-band measurements, we  use a
method similar to the traditional ``reddening-free'' {\it Q} index.

\subsection{The WFPC2 photometry. Foreground contamination}

 The observational  color -- magnitude diagrams in the F336W, F439W and F555W 
filters are shown in Figure~\ref{fig:clm}, for all stars in the 
Group 1 fields 
(open circles) and in the Group 2 fields (filled circles). 
The dashed lines indicate the magnitude  limits of the
``restricted sample'' analysed in the following sections.

As described in earlier studies a  ``blue plume'' is present, at 
$(B-V)\approx0$ (see e.g. Bianchi et al. 2001a) indicating
 H-burning massive stars, especially prominent in the Group~1 population.
 There are a few blue stars in the outer parts of NGC\,6822 (Group 2 fields). 
Most of the  blue stars in our sample 
 are found in the Group~1 fields which are closer to the 
main body of the galaxy. The majority of the blue stars is located 
in the WF3 chip of Field~36, shown in Figure~\ref{fig:f36}. 
This field contains most of the association OB15 identified 
by Hodge (1977). There are 47 blue stars with good photometry in all the 
five filters, 46 of them in the ``Group 1'' fields and only one in the
``Group 2'' fields.
Their photometry is given in Table~ 3. 

The majority of red and intermediate color stars in our sample is expected to be 
foreground contamination at the latitude of NGC~6822. 
The model of Ratnatunga \& Bachall (1985) predicts most of the foreground 
stars in the direction of NGC\,6822 with observed magnitudes 
$17 \le \mv \le 22$ to have $0.8\le (B-V) \le 1.3$, which is the 
area of the HR diagram 
where most of the stars from the ``Group~2'' fields are found. 
There is excellent agreement between the expected number of foreground stars 
in a given color and visual magnitude range, and the number of Group~2 stars
with the same colors and magnitudes, for the brighter magnitudes, 
but at  fainter magnitudes the number of stars measured exceeds the 
number of predicted foreground stars  (Figure~\ref{fig:fgs}). 
The number of foreground stars given in the figure is calculated 
from the model of  Ratnatunga \& Bachall (1985)
for an area $13 \sqamin$ (as covered by our Group~2 fields),
the  number of stars detected in each bin is also reported. 
For magnitudes above the limits of our ``restricted sample''
(next section) there is good agreement, suggesting that most stars
in ``Group~2'' are foreground, therefore we can use the ``Group~2'' sample
to estimate the foreground contamination in the ``Group~1'' sample,
above $V \approx 22$.
At fainter magnitudes, we detect significantly more objects than
predicted by the Ratnatunga \& Bachall (1985) model. However, these
authors warn that for $20>|b|>10$  or  $V>22$ their model 
should be used with  caution 
because these regions are outside the Galactic latitudes or apparent magnitudes
tested.  We cannot establish, with our present sample,  whether the 
excess of faint stars (up to 7 times more stars detected than predicted 
for (B-V)$<$0.8 and $21> V > 23$) is due to an old population 
in the outskirts of NGC~6822, or if the model of Ratnatunga \& Bachall (1985)
needs to be significantly revised in this magnitude range. The question could
be  clarified by deep imaging with a ground-based telescope. 

Using the CMD of the Group~2 sample to estimate the foreground contamination of the
Group~1 sample, we find that the majority ($\approx 70\%$) of the 
intermediate color stars ($0.8\le (B-V) \le 1.3$) are foreground objects.
The fraction drops to $\approx 30\%$ for stars with bluer 
color ($0.3\le (B-V) \le 0.8$), and is negligible at $B-V\approx0$.
We conclude that foreground stars are a significant contamination
in  the red 
part of the CMD (both from the Ratnatunga \& Bachall model and by using
 our Group~2 fields as a proxy for the foreground stars estimate) while 
they don't affect the estimate of the hot massive 
stars ($(B-V) \approx0$) content.

\subsection{Derivation of stellar parameters}

Here we use the observed photometry to determine the stellar physical 
parameters. In the first method, known as the ``Q-method'',
 we construct reddening free indexes (Q), 
using different combinations of two colors to determine the amount of
reddening and the stellar temperature concurrently.
This method is applicable to all stars that have 
good measurements in at least three photometric bands.
In the second method we determine the values of \ebv\, and \Teff\, 
using all available bands by simultaneous fitting of the observed colors to  
synthetic colors using  $\chi^{2}$ minimization. For this purpose we used 
both a modified version of the Bianchi et al. (2001a),  Romaniello (1998) 
code, and the CHORIZOS code by Ma\'{i}z-Apellaniz (2004). 
Both codes give consistent results.
This method is applicable to the stars with good  
photometry in four or more bands.

\begin{figure}
\hskip -1.cm
\epsscale{.6}
\plotone{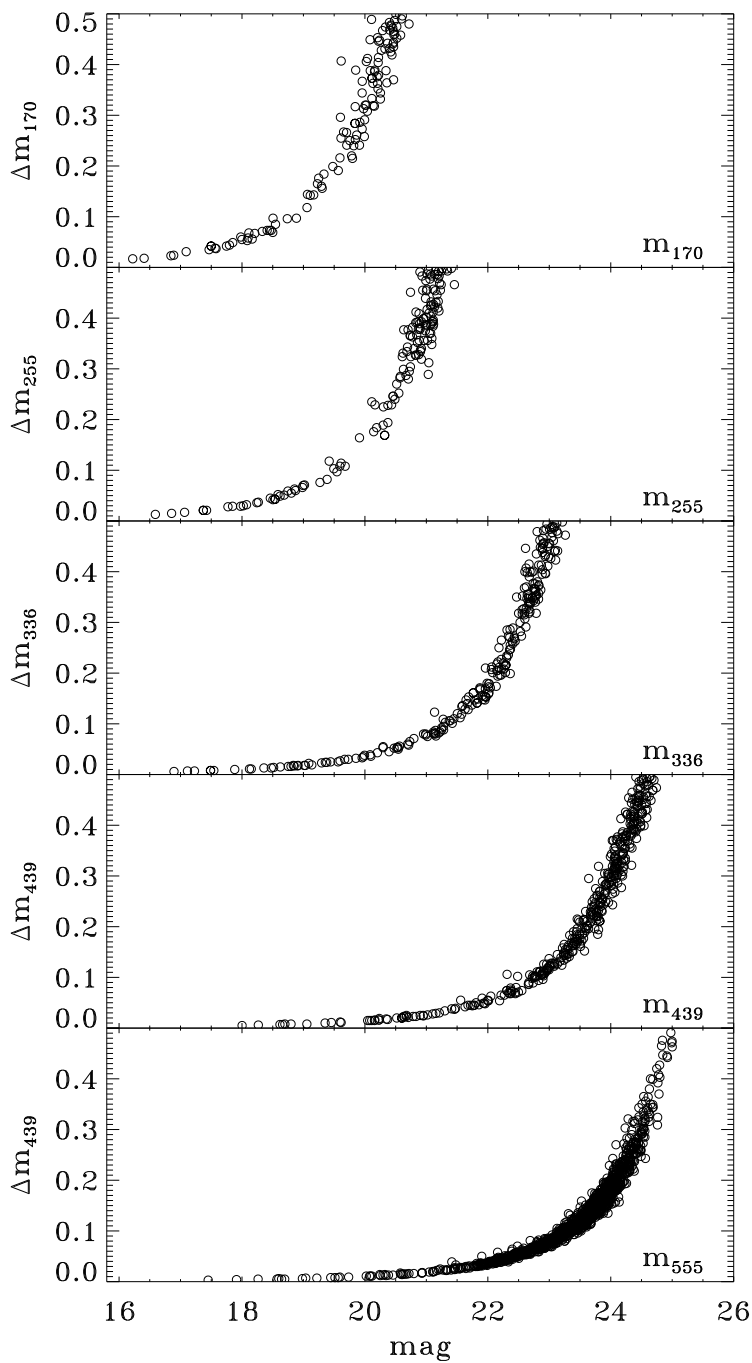} 
\caption{Error -- Magnitude diagrams for the WFPC2 photometry. The data are from a field
with long exposures; fields with shorter individual exposures mostly overlap
thus the combined magnitudes for the objects in these fields have errors comparable
to those for objects in a single field with longer exposures.  
\label{fig:err} }
\end{figure}

\begin{figure}
\hskip -1.cm
\epsscale{1.}
\plotone{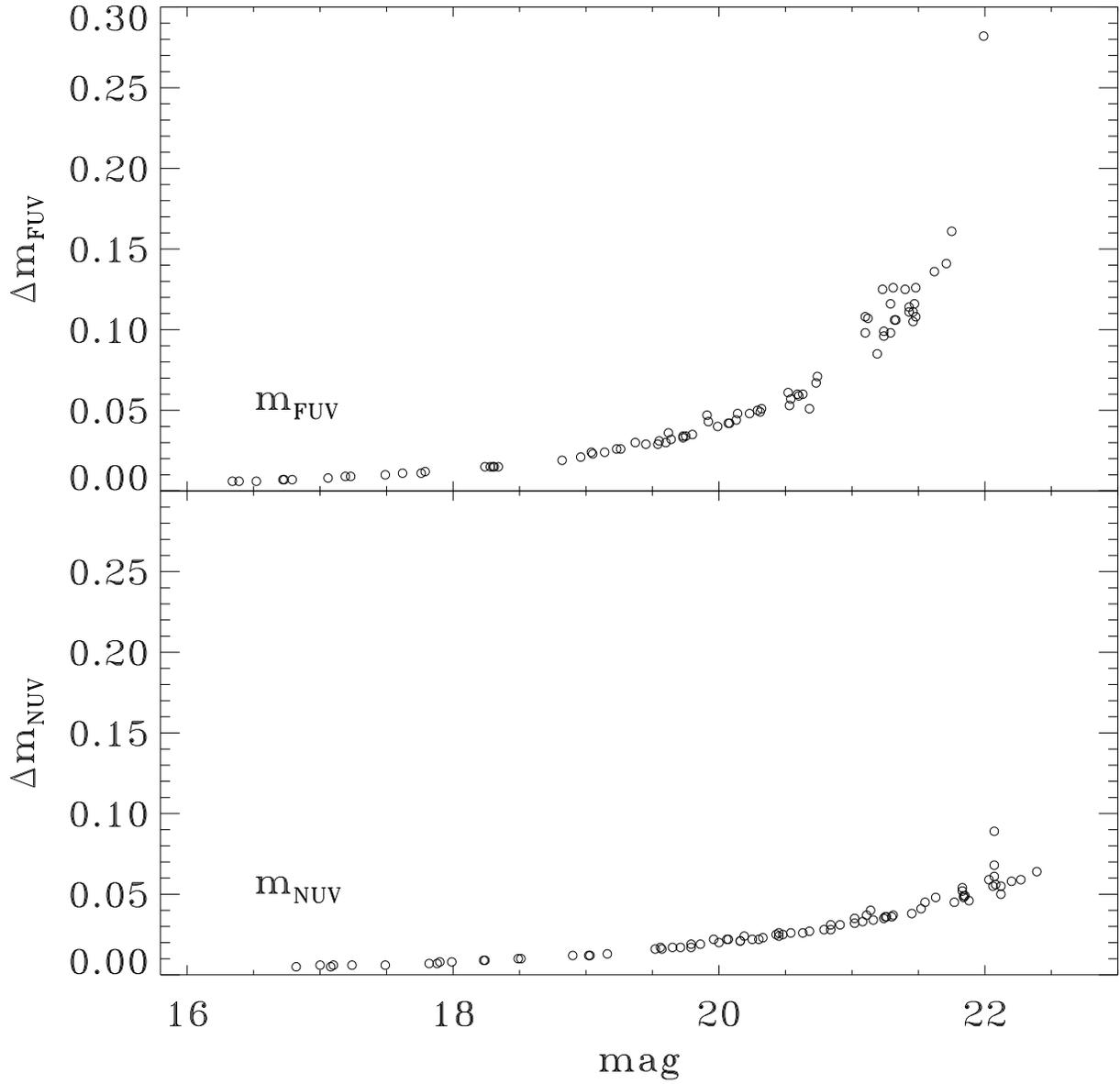}
\caption{Error -- Magnitude  diagrams for the STIS photometry.
\label{fig:st_err} }
\end{figure}

\begin{figure}
\hskip -1.cm
\epsscale{1.}
\plotone{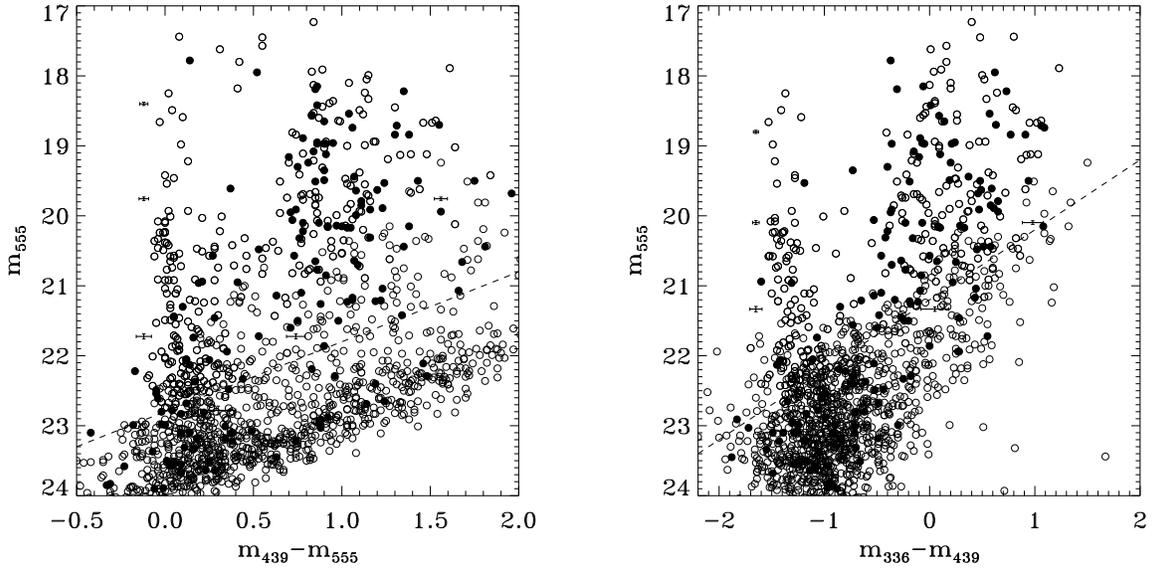}
\caption{Color -- Magnitude diagrams for stars measured in the WFPC2
fields with the F336W, F439W and F555W filters. 
Stars in the Group~1 fields are indicated
with  open circles, stars in Group~2 fields are the filled circles. 
Typical errors for representative values of magnitude and color 
 are shown. The dashed lines indicate the magnitude limits imposed
by a photometry error cut of $<$0.1, $<$ 0.08 and  and $<$0.05~ mag in 
F336W, F439W and F555W respectively, showing the  limit
of our ``restricted sample'' for the analysis described in section \ref{s_analysis}.  
\label{fig:clm} }
\end{figure}

\begin{figure}
\hskip -1.cm
\epsscale{1.}
\plotone{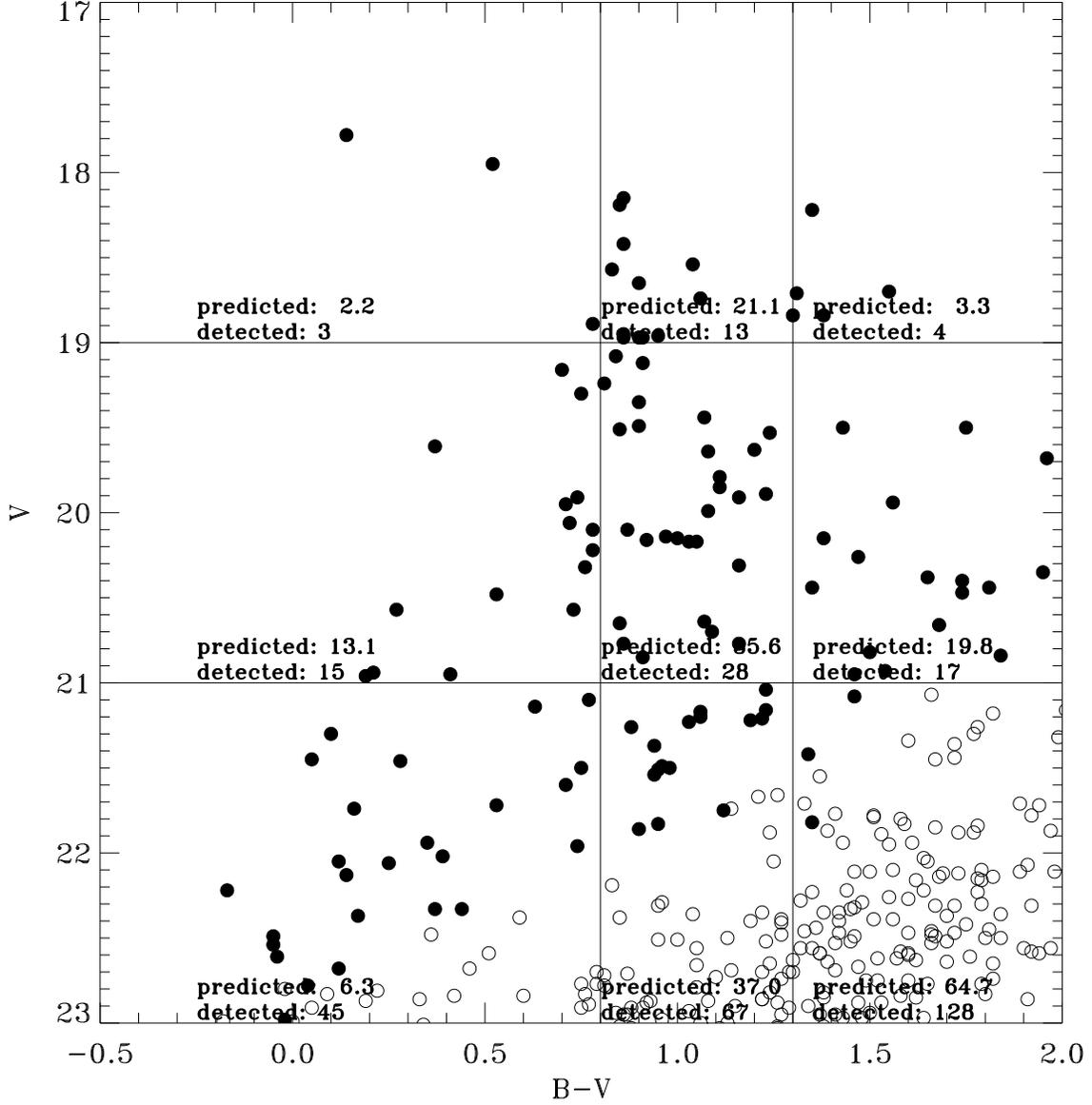}
\caption{Color -- Magnitude diagram of stars detected in the  Group~2 
fields with the number
 of foreground stars per color -- magnitude bin predicted by  the Ratnatunga 
\& Bachall (1985) model in a 13 arcmin$^2$ area. 
There is  agreement between the number of stars in Group~2 and the 
expected amount of foreground stars, except for the fainter magnitude  bins,
indicating that at the
galactocentric distance of these fields
 the stellar objects detected (brighter than the
magnitude limits of our restricted sample) 
are essentially foreground Milky Way stars. 
The excess of detected objects in the fainter magnitude bins may be
due to the uncertainty of the  Ratnatunga 
\& Bachall (1985) models for magnitudes  fainter than $V \sim 22$. 
Filled circles are the ``wider sample'' analyzed in Section 3.2. 
\label{fig:fgs} }
\end{figure}

\begin{figure}
\hskip -1.cm
\epsscale{1.}
\plotone{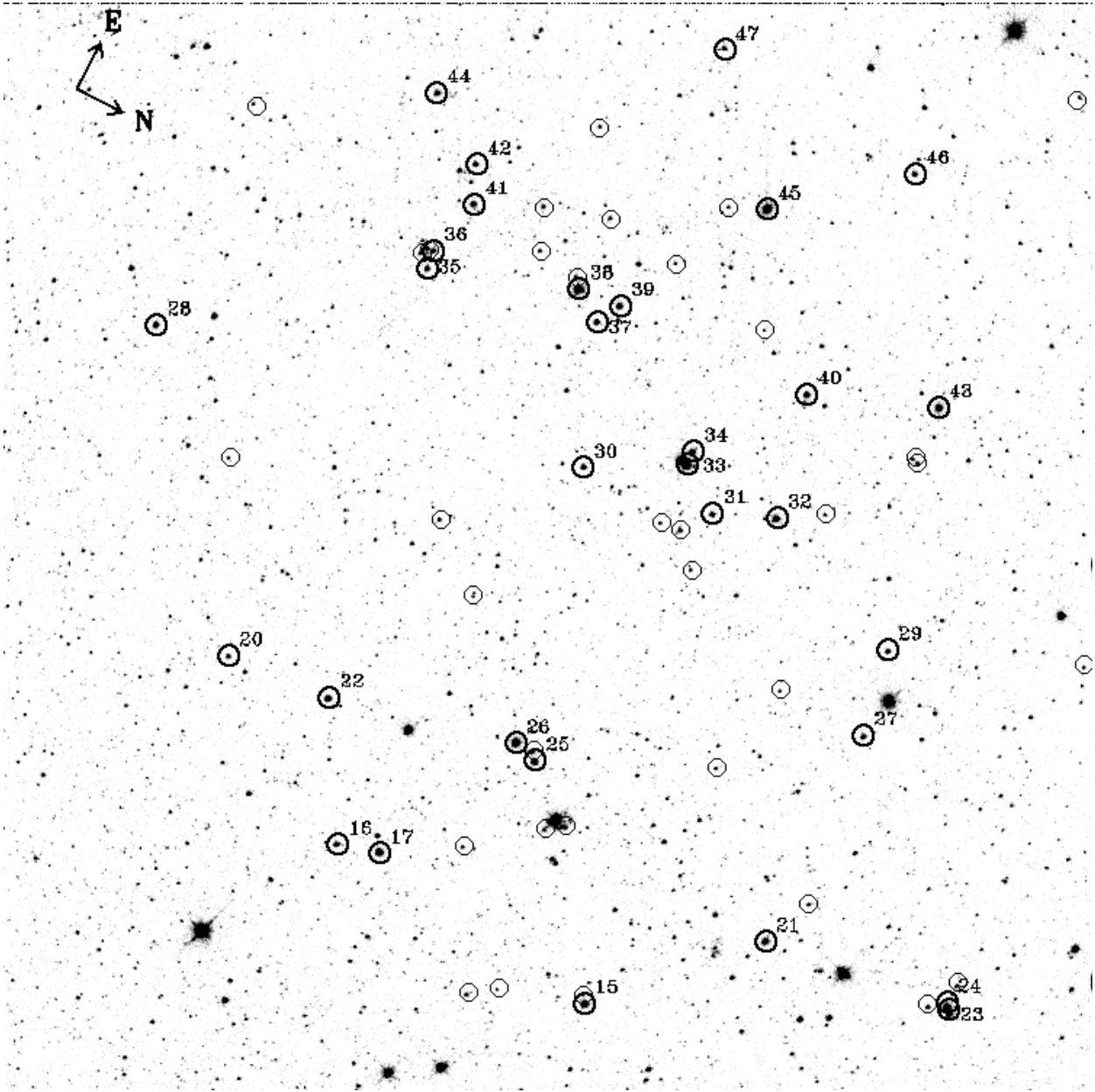}
\caption{The WFPC2 image of OB15  in the  F555W band (Field 36, WF3 chip,
80$''$ on a side). 
The stars detected in all five bands (F170W, F255W, F336W, F439W, F555W) 
are indicated with thick circles and their
identification from Table 3 
is given. 
The thin circles indicate hot stars with good photometry in only three filters
( $\magu-\magb<-1$ and $\magb-\magv<0$). 
\label{fig:f36} }
\end{figure}

\begin{figure}
\hskip -1.cm
\epsscale{.5}
\plotone{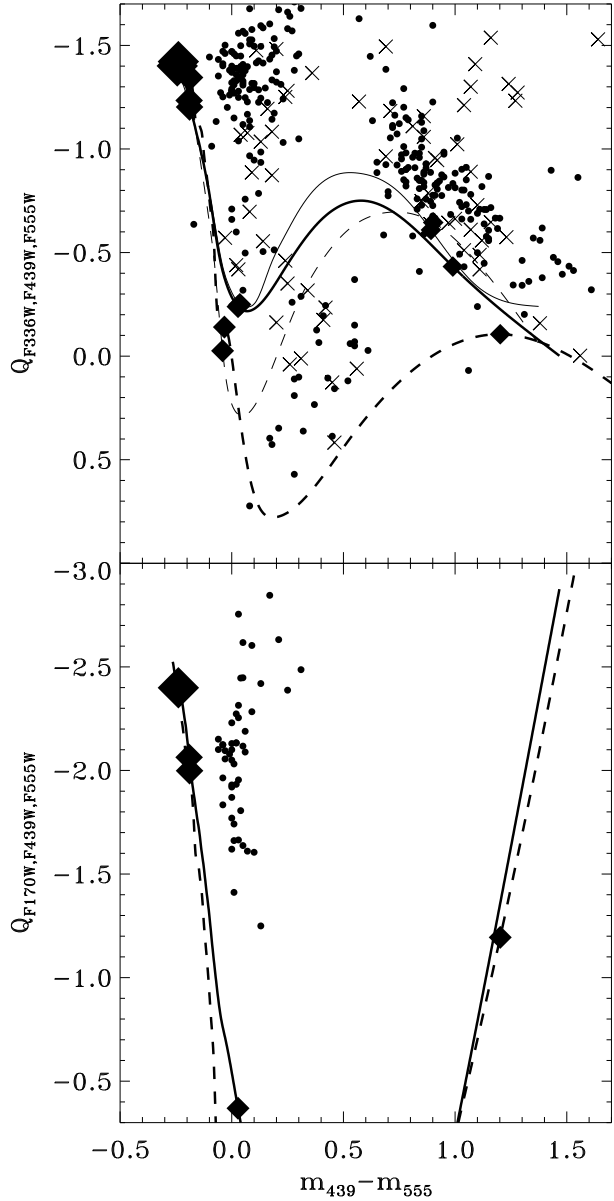}
\caption{
Upper panel: the reddening-free index $Q_{F336W,F439W,F555W}$ vs ($\magb-\magv$).
The dots represent our ``restricted sample'' and the crosses are 
the ``wider sample''.
The lines are constructed from model colors, for main sequence stars
(solid lines) and supergiants (dashed), thick lines are for solar metallicity
and thin lines for models with Z= 0.002. 
Bottom panel: $Q_{F170W,F439W,F555W}$ vs ($\magb-\magv$).
The dots represent the stars from our restricted sample that are 
detected also in the F170W band.
The model $Q_{F170W,F439W,F555W}$ (solar metallicity) 
 for main sequence and supergiant stars  are overplotted as above.
Filled diamonds of  decreasing size mark 
models of  
\Teff = 35,000 , 25000, 10000, 5000~K.  
}
\label{fig:qss} 
\end{figure}

To describe the analysis with the reddening-free index method, we first consider 
photometry in the F336W, F439W, and F555W filters, since our data reach
 fainter magnitudes in these bands (figure \ref{fig:err}).  
We limit the analysis to stars with 
photometric errors of 0.05, 0.08 and 0.10 mag in  F555W,  F439W and F336W 
respectively (``restricted sample'').
The error on the reddening-free index, and hence on the derived parameters,
is a combination of the errors in the three bands. 
We applied a progressively less stringent
error cut from the V-band to the U-band measurements, 
in order to approximately match the depth of the sample in these three bands 
for the ``blue plume'', and still have an overall good accuracy of the Q-index.
The blue plume has  average colors of  (m$_{439}$ -  m$_{555}$) $\approx$ 0.2 and 
(m$_{336}$ - m$_{439}$)  $\approx$ -1.2,  
therefore the same ``depth'' of the sample is reached for
 m$_{336}$ $\approx$ m$_{439}$ - 1.2 $\sim$  m$_{555}$ - 1. 
Our chosen  error cut of 0.1~mag for the m$_{336}$ magnitudes, 
limits the m$_{336}$ measurements to $\sim$ 21.5~mag (figure \ref{fig:err}). The
corresponding errors limits in m$_{439}$ and m$_{555}$ are 0.08 and 0.05~mag
for our ``restricted sample''.
These error limits  combined translate in a maximum uncertainty  
for reddening-free index  of Q$_{err}$$<$ 0.16, 
 which propagates into errors
on the derived parameters  according to the parameters regime, as can be 
seen in Figure \ref{fig:qss}. 
The restricted sample defined above 
is hence limited to  m$_{555}$ $\sim$ 22.5 mag, and the number of objects to 230.
For red supergiants, having  (m$_{439}$ -  m$_{555}$) $\approx$  1. and 
(m$_{336}$ - m$_{439}$)  $\approx$ 0.--0.5, the  limit 
of our restricted sample
becomes m$_{555}$ $\sim$ 20.5--21.0~mag, basically driven by the
U-band error cut of 0.1~mag. 
We also consider a less restrictive sample (``wider sample''), 
with error limits of 0.17, 0.11, and 0.07~mag in 
m$_{336}$, m$_{439}$ and m$_{555}$ 
respectively (Q$_{err}$$<$0.24), which increases the number of objects by
 130 stars, 
and extends the  limit to 
m$_{555}$ $\sim$ 23~mag. The total number of objects in the ``wider sample''
is therefore 360; values of \Teff ~ and \ebv ~ from the analysis 
described below are given in Table 3 (electronic version) for 345
stars of this sample; 
the remaining 15 objects have  colors in the range where the Q-method is  
not applicable (Q is not reddening-free), and are probably foreground
cool stars.  
Table \ref{tbl:errs} gives the number of objects in each filter within
given error limits.

The ``reddening free'' Q index is a combination of two different colors 
(usually from three bands) with the ratio of the two color-excesses,
 so that its value does not depend on the  reddening. 
The classical Q-index is constructed with magnitudes in U, B and V (e.g. Massey 1998a): 
\[Q_{UBV}=(U - B) - \frac{E(U -B)}{E(B-V)}(B - V)=\]
\[=(U - B)_0 + E(U - B)- \frac{E(U -B)}{E(B-V)}[(B - V)_0+E(B -V)]=\]
\[=(U - B)_0 - \frac{E(U -B)}{E(B-V)}(B - V)_0= Q_{UBV0},\] 

where the subscript $_0$ indicates intrinsic colors. A similar index can be
constructed for any combination of at least three filters.
The observed value of $Q$  is equal to the intrinsic one, 
hence it  provides a reddening independent
direct measurement of \Teff , 
to the extent that the ratio $C=\frac{E(U -B)}{E(B-V)}$ is actually constant. 
This is a good approximation for low values of \ebv. If we were using
monochromatic fluxes, $C$ would be constant for any amount of reddening,
and would only depend on the type of extinction, i.e. on the extinction
curve adopted. In practice, when we use broad band filters, because both  
  \Teff\, and reddening variations change the slope of
the spectrum within the band, 
$C$ is  constant only within limited ranges of  \ebv\, 
and \Teff ~ (see Bianchi \& Tolea 2006 for further discussion). 

We constructed Q indexes (as a function of stellar parameters)
in the HST filter system using a grid of synthetic colors,
obtained by applying the transmission curves of  
the WFPC2 filters to  Kurucz synthetic stellar spectra 
models of Lejeune et al. (1997) and 
Bianchi et al. (in preparation). 
Our model colors cover a large range in stellar parameters \Teff\,and log(g), for
different metallicities.
The effects of differing amounts and types of 
extinction are applied to the model spectra, 
and synthetic colors are computed
applying the filter curves to the reddened models. Therefore,
our grids of model colors allow us to also 
derive the value of {\it C} for various types of
reddening,  and to verify within what limits it remains actually constant.
We initially adopt MW type extinction ($R_V=3.1$) assuming that in our sample
most of the extinction is due to Milky Way foreground dust, 
which is confirmed by our results. 
The MW reddening law ($R_V=3.1$) was also found appropriate for NGC\,6822 by 
Massey et al. (1995). 
From our model colors, we found $C_{F336W,F439W,F555W}$ to be constant 
in the temperature range 
$\Teff \in [8000, 50000]$ and \ebv $<$ 0.7], with a value
of $C_{F336W,F439W,F555W}$ =0.96. We use the 
$Q_{F336W,F439W,F555W}$ index to derive the extinction and the stellar temperatures 
in these intervals. We can also use $Q_{F336W,F439W,F555W}$ for the stars 
in the range $\Teff \in [5000, 8000]$ and 
\ebv $<$ 0.7], with  $C_{F336W,F439W,F555W}$=0.90. 

Because a higher reddening applied to a hot star spectrum can mimic
a less reddened cooler spectrum, 
in certain parameter ranges the solution [\Teff , \ebv ] is not unique,
as can be seen from Figure \ref{fig:qss}.
 Since the Q index is 
reddening independent (within the applicable range), 
on the Color -- Q diagram 
the reddening only displaces the points in color (along the x-axis in figure \ref{fig:qss}). 
The difference in color between the observed points and the model colors
(corresponding to the intrinsic colors) 
provides an estimate of the extinction, while the value of Q provides \Teff . 
In the ranges where, for a given observed color and Q,
the intrinsic color has more than one possible value (
$Q_{F336W,F439W,F555W} \in [-0.7,-0.2]$, (m$_{439}$ -  m$_{555}$)  $>$ 0.2 
for solar metallicity, see Figure \ref{fig:qss})
different solutions for [ \Teff , \ebv ] exist. 

 In the upper panel of Figure~\ref{fig:qss} we plot the synthetic 
$Q_{F336W,F439W,F555W}$ index vs the 
color $\magb-\magv$. The observed  values are shown as 
 dots for our ``restricted'' sample (stars with error in Q better than 0.16), and with
 crosses for the ``wider'' sample (0.24$> Q_{err} >$0.16). 
The  solid lines represent model colors for  main sequence stars 
(thick: solar metallicity, thin: Z=0.002). 
The model curves show that: for  $Q_{F336W,F439W,F555W}$ $\lesssim$ - 0.7
(the exact value slightly depending on metallicity) there is only one solution;
for more positive values of $Q_{F336W,F439W,F555W}$ there are the following possibilities: 
when $(\magb-\magv) \leq 0.1$ there is a unique solution 
for [\ebv,\Teff], for stars with $0.1 <(\magb-\magv) < 0.6$ there are 
two possible solutions, and for $(\magb-\magv) > 0.6$, there are three 
possible solutions. 
In the case of this study, however, we notice that when there are 
 more than one possible solution,  the lowest
 value of \ebv~ is similar to the average \ebv ~  in the field. 
The values of  \ebv ~ from the  other possible solutions  
significantly exceed any value of \ebv ~ determined 
in these fields for the 
stars with a unique solution, and 
exceed by $\sim$ 0.8 mag the average  \ebv ~
determined in the galaxy by this and previous works. 
This difference can  be appreciated 
at a glance from figure  7. 
 Therefore we choose the solution [\ebv, \Teff\,] 
with the lowest extinction value, 
assuming that there is no reason for the extinction 
to vary extremely in such sparse fields. 
For the same reason, for those stars that have  $(\magb-\magv) > 0.5$ 
and formally a unique solution from the Q- diagram implying 
\ebv $>$ 0.5, we adopted instead the average value of \ebv = 0.27 
(as determined from the ``restricted sample'' in the
field population outside of OB15) for those stars in the ``Group 1'' fields,
and \ebv = 0.2 for those 
located in the ``Group 2'' fields.
These objects  are seen as a small ``plume'' above the model-Q 
curves for red stars in figure 7, and are indentified with flags
``c'' and ``d'' respectively in the electronic version of Table 3.
Theoretically, they could be hot stars with very high extinction,
but the majority is likely foreground stars. 

An interesting feature of the color $-$ Q diagram is the separation of
the model color curves for different gravities. 
Main sequence (solid lines), and supergiants (dashed lines)
are plotted in the \Teff\, range [50000, 4000] in  Fig.~\ref{fig:qss}. 
The curves separate for \Teff $\lesssim$ 10,000~K ($Q > -0.2$).  A 
group of $\approx$ 25 stars 
with high values of Q can be classified as supergiants, based on the 
photometry, indicating that these cooler objects  belong to
NGC~6822 and are not foreground stars. They are found in the ``Group~1'' fields, 
(4 in the area of OB15) and only 1 or 2 are found in the ``Group~2'' fields. 

We have also calculated $Q_{F336W,F439W,F555W}$ using low metallicity 
models (Z=0.002;  
thin lines in Fig.~\ref{fig:qss} upper panel) in order to 
check how  metallicity affects the color -- Q diagram. 
The solutions for hot stars (down to 10000K) do not differ significantly with 
metallicity but for low temperature stars the results for \ebv\, and \Teff\, 
 depend on metallicity.
This paper is focussed on  the hot massive stellar population of NGC6822, 
so the choice of  metallicity does not affect our results. 

For stars detected also in the  UV, we constructed additional Q indexes 
including the F170W magnitude, which 
provides a more sensitive measurement of 
the high temperatures and especially of the extinction.
As shown in the bottom panel of Figure \ref{fig:qss}, 
 $Q_{F170W,F439W,F555W}$ spans over 2~mag in
the \Teff~ range 10000-35000~K, compared to the $\sim$1~mag
variation of Q$_{F336W,F439W,F555W}$. However, 
the larger uncertainties in the UV photometry, due to the 
lower sensitivity of the CCD at UV wavelengths, limit the advantage 
of these broader wavelength coverage to the hottest objects.
We found, from our model colors,  $C_{F170W,F439W,F555W}$=4.15,
 approximately  constant for 
 $\Teff \in [11000, 50000]$ and  $\ebv <$ 0.7.
The model values of  $Q_{F170W,F439W,F555W}$ for
 low gravity stars again separate from the dwarfs for temperatures between
10000K and 8000K. This part of the diagram 
is not shown since only the hottest stars 
in our sample are detected  in F170W. 
 The analysis using 5 band photometry is restricted to objects with  
photometric accuracy in F170W and F255W  better than $0^{m}.2$.
There are 47 such stars in the WFPC2 sample.
All of the stars with good photometry 
in five bands lay in the unique-solution part 
of the Color\,--\,Q diagrams, and the values of \ebv\, and \Teff~
 obtained from different
 Q indexes are consistent within the errors. 

For our  restricted sample \ebv\, varies between 0.17 and 0.37,
with a mean value of \ebv\, of 0.27 
 which is in agreement with the estimate 
of Massey et al. (1995) for the outer parts of NGC\,6822. 

\begin{figure}
\hskip -1.cm
\epsscale{.6}
\plotone{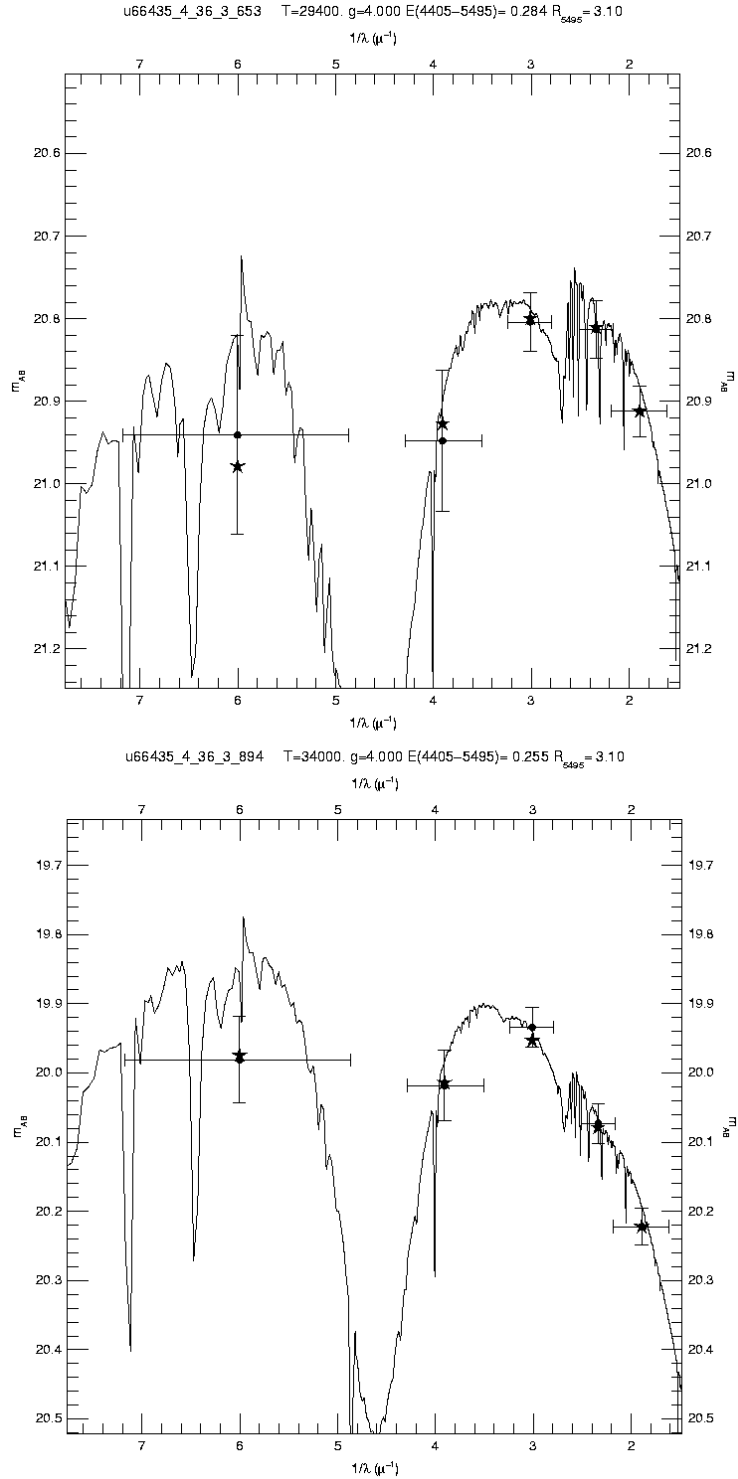}
\caption{Two examples of $\chi^2$ fits obtained with the CHORIZOS code. 
The crosses are the observed magnitudes with their errors, the 
stars are the synthetic magnitudes with the width of the filters.
The model spectrum from which the synthetic magnitudes are generated is also shown.
The lines appear very strong in the spectrum because of the flux scale in magnitudes.
\label{fig:cho} }
\end{figure}

In the second method we derive the values of extinction and stellar temperature by 
simultaneous $\chi^{2}$ minimization fitting of all the observed colors with 
the same library of model colors.  The advantage of this method is 
the use of  all the available photometric information simultaneously,
each magnitude being weighted according to its uncertainty.
We used the CHORIZOS code 
 by Ma\'{i}z-Apellaniz (2004),
as well as a code developed by us (based on the method by Romaniello 1998) 
 and find consistent results. 
We obtain a good fit with $\chi^{2} \in [0.6,5.5]$ for 
approximately $60\%$ of the stars with five band photometry. Two examples of
 good fits are shown in Fig.~\ref{fig:cho}. The model 
spectrum that best  fits the observed colors is shown along with the model magnitudes
 and the observed magnitudes with their errors.  The derived \Teff\, and \ebv~
 correlate well with the values obtained from the Q -- method within the errors,
although there is a slight trend of higher extinction values, and consequently
\Teff , from the $\chi^{2}$ fit. 

In both the reddening-free and the $\chi$$^2$ fitting method we used 
MW-type extinction with R$_V$=3.1 and also tried UV-steeper extinction
laws, such as LMC-type extinction. We found the MW-type extinction to
provide better match to the observed photometry, consistent with the
result that most of the extinction is due to foreground MW dust.  
In order to construct physical HR diagrams (next section), 
 for the stars with five band photometry we adopt the values 
of \Teff\, and reddening from the   $\chi^{2}$ fit (given in the printed
version of Table 3)  and 
 for the stars with three good measurements
 the values from the Q -- method. Stars with B and V measurements
having errors better than 0.2~mag, but errors larger than our ``wider
sample'' limits in other bands, are also included the end of Table 3
(electronic version). 

\subsection{STIS photometry}
\label{s_stis_analysis}

Most of the UV sources (72 out of 80) in the STIS imaging 
of OB~8 have measured counterparts in the WFPC2 photometry 
of Bianchi et al. (2001a) that provides additional measurements in 
four bands: F255W, F336W, F439W, and F555W. 
We derived temperatures and \ebv\, by comparing the photometry to the 
model colors as described in the previous section,
using $\chi^2$ fitting. 

The additional STIS FUV band gives us the opportunity to refine the determination of  
extinction and temperature with respect to our previous work.
The errors in the WFPC2 imaging of Bianchi et al. (2001a) in the
 F255W and F336W filters are larger than 0.2~mag for more than one third of the
 sample,  and our
 new STIS UV photometry with relatively small errors significantly 
improves the determination of the stellar parameters. 
The stellar temperatures determined from the  $\chi^2$ fit including UV bands 
tend to be higher than the previous results.  
The HR diagram of the Hubble V region is shown in fig. ~\ref{fig:hrs} with 
filled circles. The results from the 
photometry of Bianchi et al. (2001a) are shown with  crosses for comparison.
 The photometric measurements and derived stellar parameters, as well as the identification
with the sources from Bianchi et al. (2001)  are listed 
in Table 4. The coordinates are from the STIS images, however
we applied a constant shift of $+0.684''$ and $-0.54''$ in R.A.,
 and $-0.684''$ and $-0.72''$  in declination, for fields o66410 and o66420 respectively, 
in order to
register the astrometry to the WFPC2 coordinates of Bianchi et al. (2001). 
After the shift, positions for matched stars coincide within 0.2'' arcseconds.

\section{Results and Discussion}

\begin{figure}
\hskip -1.cm
\epsscale{1.}
\plotone{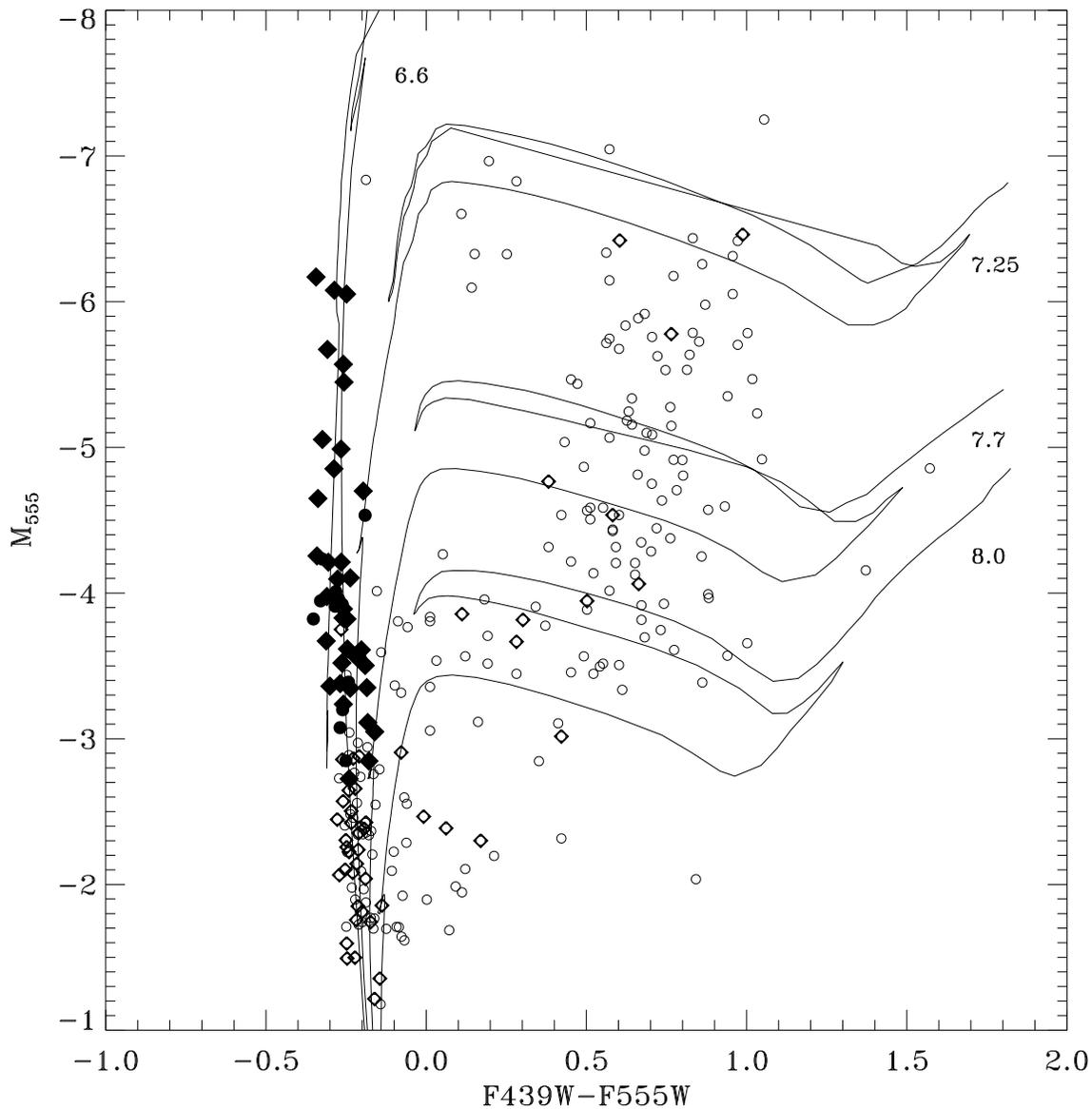}
\caption{HR diagram for the stars in the Group 1 fields (``wider sample''). 
The observed magnitudes and colors have been dereddened as described in section \ref{s_analysis}.
The filled symbols represent the stars with good photometry in all 
the five filters used. Stars in the region of the OB15 association are shown as 
diamonds, stars outside this region with circles.  
 Isochrones for Z=0.004 from
 Girardi et al. (2002) 
and Bertelli et al. (1994) are shown, with ages (in log years) indicated.
\label{fig:cmp} }
\end{figure}

\begin{figure}
\hskip -1.cm
\epsscale{.6}
\plotone{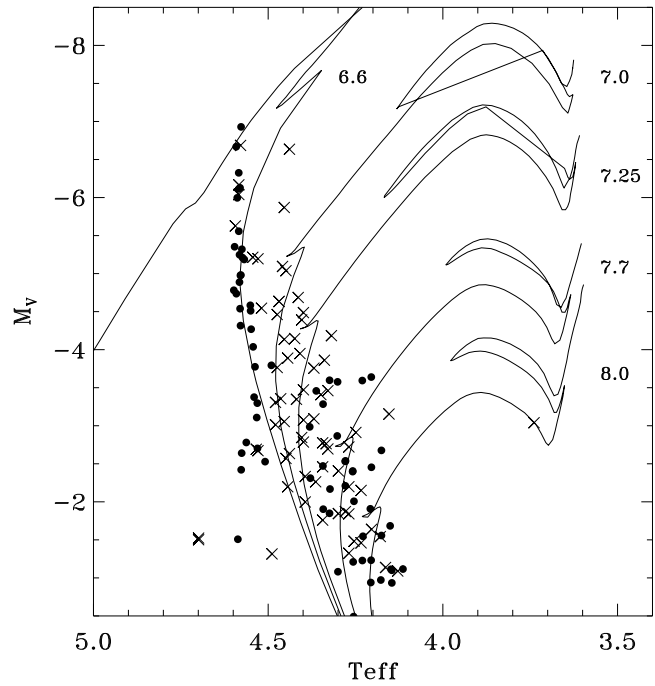}
\caption{The HR diagram of the Hubble V region: The results from combined STIS
(this paper) and  
WFPC2 (Bianchi et al. 2001a)  photometry 
are shown with filled circles. For comparison, 
the results from Bianchi et al. (2001a) are also plotted (crosses).
 Isochrones for Z=0.004 from
 Girardi et al. (2002) 
and Bertelli et al. (1994) are plotted,  ages (in log years) are indicated.
\label{fig:hrs} }
\end{figure}

Absolute magnitudes for our sample  were calculated using the reddening derived in the 
previous section and a distance modulus of DM=23.47 (McGonegal at al. 1983).
 In Figure~\ref{fig:cmp} the Color--Magnitude diagram for stars in the Group 1 
fields is shown with superimposed theoretical isochrones for Z=0.004 from
 Girardi et al. (2002)
and Bertelli et al. (1994). Isochrones for solar metallicity do not fit
as well the observations. 
The majority of the Group 1 stars with five band photometry
(34 stars, or 74\%) are 
in the association OB15 (Hodge 1977) centered at 
$\alpha=19^h\,45^m\,14^s$ , $\delta=-14^{\circ}\,45^m\,7^s$, 
with an angular size of $\approx 90''$ ($\approx 200pc$ linear size),
and a distance from the center of the galaxy of $\approx 5'$ or 730 pc. 
The main part of the association is included 
in the WF3 chip of Field~36 shown in Fig.~\ref{fig:f36}.
For these stars we estimated \Teff $>$ 20,000~K (spectral type 
earlier than B2). Their 
``scale distance'', defined as the largest distance 
between two close neighbors 
in the group 
(see Battinelli 1991 and Ivanov 1996) is $d_S=26\arcsec \, (46 pc)$, 
and the mean distance between close neighbors is $d_M=5\arcsec \, (13 pc)$, 
with $\sigma=4.5\arcsec \,(11.4 pc)$.
If we add fainter hot star candidates that have photometry only in three bands, 
using the $(\magu-\magb)_o <-1$ and $(\magb-\magv)_o <0.0$ criteria to select 
 stars hotter than 16,000~K (i.e. earlier than sp.type $\sim$ B5~V), we find
35 additional stars in OB~15 (down to magnitude V$_o$= 22.2 (V$\approx$ 23.0, M$_V$=-1.27)
and the association 
properties become  $d_S=13\arcsec \,(32 pc)$ and $d_M=4\arcsec \,(10 pc)$  with 
$\sigma=3\arcsec\, (8 pc)$. The total number of stars with estimated 
\Teff $>$ 16,000~K in OB~15, is therefore about 70. 
 The expected foreground contamination is negligible for the hot star sample 
(Section 3.1), 
so we assume all of the observed massive stars to be members of the OB association. 
The HR diagram indicates the stellar population in OB15 to be  younger than 10 Myr.

Our data, geared towards the characterization of the young massive
stars, do not provide conclusive  information on the number of red supergiants. In fact,
the U-band filter drives the magnitude limit in our ``restricted sample''
to very bright limits for cool stars 
($m_{555}$ $\approx$ 20.5, or $(m_{555})_o$ =19.7/19.3 in OB8 and OB15
respectively).  If we consider the (B-V) color only, 
we can reach fainter magnitudes (see figure 4) but we have no way to
separate the foreground (MW halo) dwarfs from the  NGC~6822 supergiants, whose
apparent magnitudes are in the same range (e.g. Massey 1998c, Bianchi et al. 2001b).
The number of red foreground stars ($(B-V)$ $>$ 1.3 ) expected from the model 
of Ratnatunga \& Bachall (1985) is 0.35 per square arcmin down to V=19 and 
1.5 per square arcmin between V=19 and 21. Therefore, we expect 24 and 35
red foreground stars brighter than V=21 in our Group~2 and Group~1 samples
respectively. The  number of stars with measured photometry $(B-V)$ $>$ 1.3, V $<$ 21,
is 14 and 32  respectively in the Group 2 and Group 1 samples.
Considering the effects of small number statistics, no detection of red 
supergiants can be claimed without spectroscopic follow-up, which we plan 
to pursue.  
A better estimate of the evolved stellar population in this region can 
be obtained from the red 
supergiants survey of Massey (1998c). This survey is complete 
(photometrically) to V=20.5.
The OB15 association is contained in 
 Massey's N6822-C field. There is only one 
star (NGC\,6822c-438) from the Massey (1998c) catalog inside the  OB15 area, 
and it is classified by the author as a foreground object. The 4 closest neighbors are 
also classified as foreground stars, so we can conclude that there are no red supergiants 
in the region of OB15, and we are looking at stars born in a very recent star 
formation process in that field.   The non-detection is consistent with
the expected number from evolutionary models. 
In the initial-mass range where both blue and red supergiants are
expected to cohexist in the HRD, 15 to 30 \Msun 
(e.g. Salasnich, Bressan \& Chiosi 1999), our sample is 
complete only for the blue objects. At the higher end of this mass range, 
even if all of our  blue stars were supergiants,  
 we would expect less than 1 red supergiant 
in the sample (although the ratio depends on metallicity, and rotation:
 extremely high rotation
can reverse the ratio, see Maeder \& Meynet 2001).  

\begin{figure}
\hskip -1.cm
\epsscale{.8}
\plotone{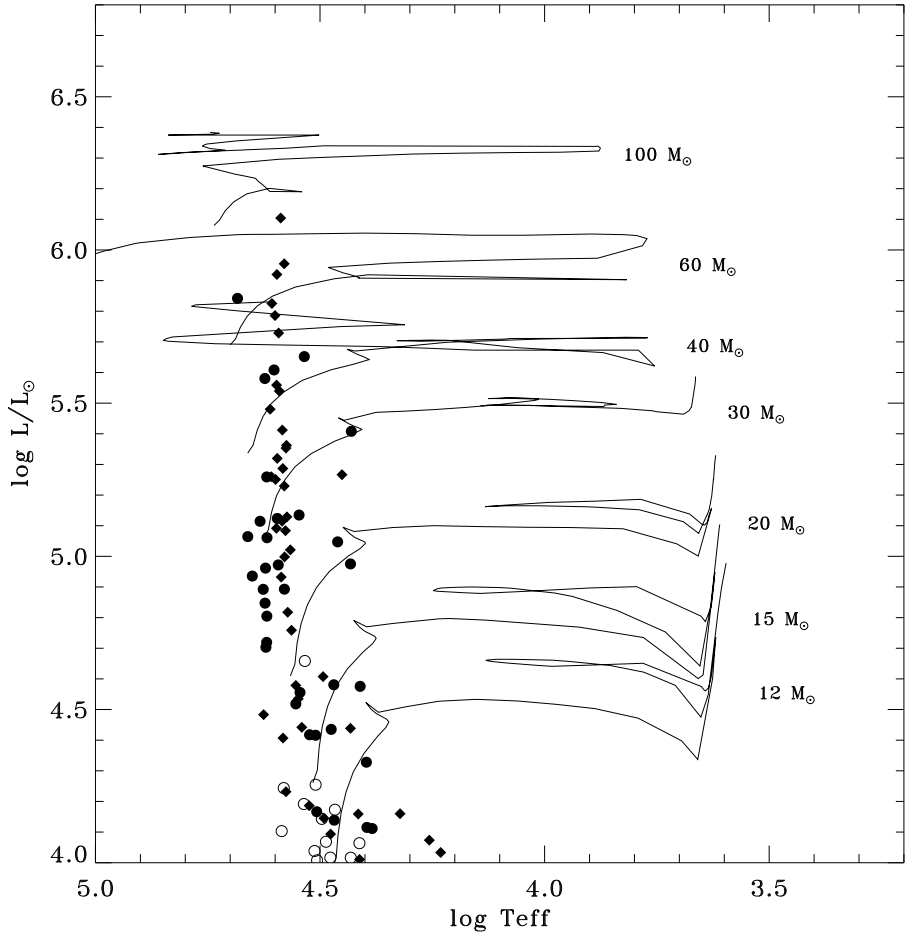}
\caption{The HR diagram of the hot stars 
in the area of the OB15 association 
(filled circles are stars with 5-band photometry, empty
circles are stars with U,B,V only) 
and in the OB8 field (diamonds).
 Overplotted are the evolutionary tracks of Fagotto et al. (1994) for Z=0.004. 
Redder stars with the same range of luminosity are not shown because
the foreground contamination is significant. 
\label{fig:mas} }
\end{figure}

The physical HR diagram for the associations OB~15 and  OB8 is   
shown in Fig.~\ref{fig:mas} with  the evolutionary tracks 
of Fagotto et. al. (1994) for Z=0.004. 
The initial-mass value is indicated for each track. 
The lower mass limit ($\approx$ 12 \Msun) reflects the depth of our photometry
for stars with 5-band measurements: 
the magnitude limit is \magv=21.5, which 
for an average  \ebv\,=0.27 and DM=23.47  corresponds to 
M$_V$=-2.7 or spectral type $\sim$ B2V.
A few very massive star candidates are found. The errors on  photometrically
derived luminosity and mass are unavoidably large, and will be refined with follow-up
spectroscopy.  The most massive star in OB~15, 
taking into account the uncertainties in  temperature and luminosity, 
can be placed within a range of masses between 40 and 70 \Msun. The range
is slightly higher for OB8. 
For the objects with good photometry in all five bands the uncertainties 
in the photometrically 
derived \Teff\, are typically $\sim$ 20\%. To refine the stellar parameters and to determine more
 precisely the mass distribution, a spectroscopic study needs to be done.
The UV bands F170W, and F255W used in this study give us an opportunity to
 distinguish between the optically  brightest (\MV), and the 
bolometrically most luminous stars with
 highest effective temperatures and largest bolometric corrections. 

The association OB8  contained in our STIS observations is much more compact 
(size about 35\arcsec or 85pc, of which 25x25$\as$ or 60x60pc are
covered by the STIS image) and 
densely populated. We detect 80 stars in our STIS FUV and NUV imaging (25x25 \as) of this region, 
down to a limit of  m$_{NUV}$=22.5 (V$\approx$ 23.4), and 
derive physical parameters for 72 of them
that are also contained in the WFPC2 photometric sample of Bianchi et al. (2001a).
The mean distance between closest neighbors is 
 $d_M= 1.2\arcsec (3~pc)$, and the maximum
separation (``scale distance'') is $d_S=7.5 \arcsec (18~pc)$. 
There is  an HII region associated with this object, Hubble~V. Its nebular properties
were studied by O'Dell et al. (1999), and its young stellar population 
by Bianchi et al. (2001a).
The HRD derived from our STIS photometry is shown in Fig.~\ref{fig:hrs}, 
together with the previous results from Bianchi et al. (2001a). 
The  temperatures estimated for most stars in this sample 
are $\Teff > 15000K$, consistent 
with spectral classes earlier than $\sim$B5V.
The extinction is higher in this region, 
having a mean value \ebv\,$\sim$0.4, with large variations.

The values of \ebv~ found in our WFPC2 fields suggest that in regions
far from the galaxy center the  reddening is mostly Milky Way foreground dust. 
The average value for all the stars in the Group~1 fields is \ebv\,=0.27 with a 
standard  deviation of 0.14. 
For the hot stars sample the mean value of the reddening obtained by  
$\chi^2$ fitting is \ebv=0.29 with $\sigma=0.06$.
The stars in the Group~2 fields show a similar 
distribution of reddening 
with a mean value of \ebv\,=0.22  and standard deviation of 0.12. 
The large scatter and somewhat lower value than the average
extinction towards NGC~6822 is likely due to the fact that most
 stars detected  in this field are foreground MW stars. 

\section{Conclusions}
\label{sconcl}

We have explored with HST 
multi-band imaging four extremely different environments
in NGC~6822. In the ``Group~2'' fields, $\approx$ 1.5-2.0~Kpc
 off (North of) the galaxy center, we 
find a very small number of hot stars,
comparable to the expected  number of foreground stars. 
The entire sample in the ``Group 2'' fields is dominated by foregound stars, 
at least down to $V \sim 22$, and 
consistently, we find in this area
the smallest extinction (average \ebv\,=0.22) measured in this galaxy. 
At fainter magnitudes, we find up to $\sim$7 times more stars than
the Ratnatunga \& Bachall (1985) model for MW stars predicts. 
However, these authors warn that their extrapolation to faint magnitudes is
uncertain, so our fainter objects may either be stars in the outskirts
of NGC~6822  or foreground stars that the current MW model fails to predict.  
The ``Group 1'' fields, covering an area of 19 \sqamin ~ ($\approx$ 402~kpc$^2$)
East of the galaxy's main bar, include the OB15 association ($\approx$ 90\as ,
or 200pc size). In OB15 there are  34 (out of the 46 total in the ``Group 1'' sample) 
hot massive stars hotter than $\sim$ 20,000K
(in our ``restricted sample'' photometry with limit V$\approx$21.5,
 corresponding to a $\sim$B2V star at the distance of NGC~6822).
About 35 additional stars (detected in $U,B,V$ only)
have an estimated \Teff $>$ 16000~K.
The HRD indicates an age of a few million years for OB15. 
The general field population, in the ``Group 1'' fields excluding OB15,
contains much fewer hot massive stars.
We  found twenty-five A-type supergiant candidates in 
the Group~1 fields outside of OB~15. These objects are visually bright and 
would be ideal targets for ground-based spectroscopy to determine chemical
abundances.

The fourth environment, studied with STIS imaging, covers a much smaller
and crowded area, including most of
 the OB8 association in the HII region Hubble V, whose
 nebular properties and stellar content were 
previously studied by O'Dell, Hodge \& Kennicutt (1999) and by Bianchi et al. (2001a)
respectively. Our new STIS imaging provides a gain of $\approx$2 mag in NUV over
our previous study (Bianchi et al. 2001a), 
and the first FUV  imaging, as well as the best resolution (0.06 pc/pxl) view, of this compact
cluster. 
In this $25''\times25''$ [60 $\times$ 60~pc] field, 
we measure 80 stars in both NUV and FUV, down to magnitude 22.5 in 
NUV 
 (V$_o$ $\approx$ 22.2, $\approx$B5V). 

To compare the general properties of the young stellar populations in 
OB15 and OB8, we estimate the average 
density of hot stars 
per unit area, 
down to the magnitude limit of
V$_o$ = 22.2 or M$_V$=-1.27 (approximately spectral type B5V). 
We find a density of 40 and 460 hot stars/arcmin$^2$, i.e. 
0.0019 stars/pc$^2$  and  0.0218 stars/pc$^2$ in OB15 and OB8 respectively,
adopting sizes of 80$''$ and  25$''$ for OB15 and OB8.
In the general field (``Group~1'') the density of hot stars detected 
is much lower, 0.7  /arcmin$^2$ (3.x10$^{-5}$/pc$^2$).
Assuming that the IMF does not vary, 
and given the similar ages of these two associations 
as indicated by their HRDs, the ratio of hot stars density above the 
 same intrinsic magnitude limit should correspond directly to the ratio
of star formation {\it density} among the two regions. Therefore, 
the SF surface density is  higher in OB8 than in OB15 by a factor of 12,
while the total SFR is the same in both associations (approximately 70  stars
hotter than 16,000~K).
We scaled the total number of stars with 
M $>$ 12\Msun~ according to their absolute luminosity (42 and 44 stars),
to derive the total mass of the clusters. 
Assuming a Salpeter IMF ($\alpha$=2.35) over a mass range of 100$-$1  \Msun~
(for comparison to other works on Local Group galaxies)
we estimate a cluster mass  of M$_{tot}$$\sim$3.5-4 x 10$^3$ \Msun~.
By using instead
$IMF=AM^{-\alpha}$ with $\alpha$=2.3 in the range 0.5-100\Msun~ and $\alpha$=1.3 
in the range  0.1-0.5\Msun~ (Kroupa  et al. 2001) 
we estimate a total mass of the clusters
of M$_{tot}$$\sim$6-7 x 10$^3$ \Msun~ in the wider mass range of 0.1-100\Msun.

This result is interesting in view of the numerous current studies
estimating the SF in more distant galaxies, or in nearby extended  galaxies
with lower resolution imaging (e.g. Calzetti et al. 2005, Bianchi et al. 2005, 
Thilker et al. 2005 
and references therein). In the case of the two star-forming
regions studied, integrated measurements such as UV or IR emission
might estimate the total SF to be equal, but not discern the 
extremly differing spatial properties.  The H$\alpha$ emission, another indicator
of star formation, also reflects the different spatial properties of the 
associations. We measured the H$\alpha$ emission in the regions of H~V and OB15
using publicly released images from the recent NOAO survey of the Local Group
(Massey et al., 2006), and examined the H$\alpha$ morphology. 
The H$\alpha$ flux, when integrated over the spatial extent of the 
stellar associations, is about 3 times higher in H~V than in OB15. 
However, while the  H$\alpha$ emission in H~V appears compact, as is the
stellar cluster, the OB15 association is surrounded by a larger,
multi-shell like  H$\alpha$ envelope. When we integrate the  H$\alpha$ flux over this
larger bubble, seemingly associated with the OB~15 stars from its morphology,
its total flux is comparable to the emission from H~V. 
It is also important that we could extend the census of massive stars in these
two associations down to early B spectral types, where a few Myr age difference
would not change the sample (numerical) statistics, making the
comparison robust.  Earlier ground-based
surveys comparing the massive star content of Local Group galaxies
were necessarily limited to higher masses, 
where both small number (IMF) statistics, and a few Myrs difference in the 
ages of the very young clusters (dominating the census of massive stars)
can bias the results significantly.  
 For example,  Massey et al. 1995 seminal paper 
on three representative Local Group galaxies M31, M33 and NGC6822, 
compared  the number of stars more massive than 40\Msun~ per Kpc$^2$.
From the massive star content, we estimated the
 total mass of the associations, and found   $\sim$ 4 ~x~10$^3$\Msun~
for a mass range of 1-100\Msun (Salpeter IMF), or   $\sim$ 7~x~10$^3$\Msun~
extending the mass range to 0.1 \Msun~ with the IMF of Kroupa (2001).

Our results confirm and quantify previous evidence that the recent/current
star formation in NGC~6822 occurs episodically with extreme spatial variations
in intensity and modality. 
The earlier star formation in this galaxy, according to  Gallart et al. (1999) and 
Wyder (2003) has proceeded smoothly (relatively constant, or
slightly increasing with time) until $\approx$ 1 Gyr ago. 
Of course, while time and space variations of the recent star formation can be
appreciated in detail from the study of young massive stars, such 
distinctions cannot be made for older populations, where the SFH is 
inferred from an HRD integrated over longer epochs, and dynamical relaxation
blurs or erases initial spatial stuctures.   
 More extended imaging studies of the entire galaxy are planned  
 using GALEX (far-UV and near-UV) and ground-based imaging. 
Follow-up detailed studies of the stellar properties are planned with ground-based
(VLT) spectroscopy.

\begin{acknowledgments}
Acknowledgments: We thank the referee for insightful comments,
and J. Maiz for assistance with installation and use of the 
CHORIZOS code.
 This work is based on data from the Hubble Space Telescope.
Support for program GO8675 was provided  by NASA  through a grant
from the Space Telescope Science Institute, which is operated by the
Association of Universities for Research in Astronomy, Inc., under
NASA contract NAS5-26555.  The work was also partly supported by NASA grant 
NAG5-9219 (NRA-99-01-LTSA-029).

\end{acknowledgments}

\renewcommand{\arraystretch}{.6}
{\rotate
{\footnotesize
\begin{deluxetable}{lccccccc}
\tabletypesize{\tiny}
\tablewidth{0pt}
\tablecaption{The HST data  (program GO-8675) 
\label{tbl:flds}}
\tablehead{
\\
\multicolumn{1}{l}{Field (Dataset root)}  & \multicolumn{1}{c}{RA (J2000)} &  \multicolumn{1}{c}{Dec (J2000)}   &
\multicolumn{5}{c}{Exposure Time [s] in each filter } \\
}
\startdata
\\
\multicolumn{1}{l}{}  & \multicolumn{1}{c}{} & \multicolumn{1}{c}{}  & 
\multicolumn{5}{c}{ ------ WFPC2 imaging --------}   \\
\multicolumn{5}{c}{}   \\
\multicolumn{1}{l}{}  & \multicolumn{1}{c}{} & \multicolumn{1}{c}{}  & 
\multicolumn{1}{c}{ F170W}  & \multicolumn{1}{c}{ F255W}  & \multicolumn{1}{c}{ F336W} & 
\multicolumn{1}{c}{ F439W} & \multicolumn{1}{c}{ F555W}  \\
NGC6822-PARFIELD10 (u6641) &  19 45 06.47 &  -14 47 05.8 &  $2\times300$ &  $3\times400$ &  &  &  \\
 & & & 					$2\times500$ &  $2\times500$ & $2\times350$ & $2\times350$ & $2\times100$ \\
 & & &					$2\times700$ & & & & \\ 
NGC6822-PARFIELD20 (u6642) & 19 44 46.07 &  -14 38 17.2 &  $2\times300$ &  $1\times300$ &  &  &  \\
 & & & 					$2\times500$ &  $2\times400$ & $2\times350$ & $2\times350$ & $2\times100$ \\
 & & &					$2\times700$ &  $2\times500$ & & & \\ 
NGC6822-PARFIELD30 (u66430) & 19 44 47.97 &  -14 38 13.2 & --- & --- &  $2\times260$ & $2\times260$ & $2\times100$ \\
NGC6822-PARFIELD31 (u66431) & 19 45 12.62 &  -14 42 45.8 &  $2\times500$ &  $2\times500$ & $2\times300$ & $2\times260$ & $2\times100$ \\
NGC6822-PARFIELD32 (u66432) & 19 45 14.25 &  -14 42 36.9 &  $2\times500$ &  $2\times500$ & $2\times300$ & $2\times260$ & $2\times100$ \\ 
NGC6822-PARFIELD33 (u66433) & 19 44 40.29 &  -14 38 59.5 &  $2\times500$ &  $2\times500$ & $2\times300$ & $2\times260$ & $2\times100$ \\ 
NGC6822-PARFIELD34 (u66434) & 19 44 39.18 &  -14 39 14.1 &  $2\times400$ &  $2\times350$ & --- & --- & --- \\
NGC6822-PARFIELD35 (u66435) & 19 45 12.71 &  -14 44 23.9 &  ---   &  ---   & $2\times260$ & $2\times260$ & $2\times100$ \\
NGC6822-PARFIELD36 (u66436) & 19 45 11.26 &  -14 45 30.2 &  $2\times500$ &  $2\times500$ & $2\times300$ & $2\times260$ & $2\times100$ \\ 
NGC6822-PARFIELD37 (u66437) & 19 44 37.87 &  -14 39 22.5 & $2\times500$ &  $2\times500$ & $2\times300$ & $2\times260$ & $2\times100$ \\  
NGC6822-PARFIELD38 (u66438) & 19 44 44.88 &  -14 38 22.1 & $2\times500$ &  $2\times500$ & $2\times300$ & $2\times260$ & $2\times100$ \\ 
 & &   &  &   &  &  &  \\  
\multicolumn{1}{l}{}  & \multicolumn{1}{c}{} & \multicolumn{1}{c}{}  & 
\multicolumn{5}{c}{ ------ STIS imaging --------}   \\
\multicolumn{5}{c}{}   \\
\multicolumn{1}{l}{}  & \multicolumn{1}{c}{} & \multicolumn{1}{c}{}  & 
\multicolumn{2}{c}{FUV-MAMA}  & \multicolumn{3}{c}{NUV-MAMA} 
\\
NGC6822-HV-LBFLD1  (o6641) & 19 44 52.20  & -14 43 14.5 & \multicolumn{2}{c}{1320.} & \multicolumn{3}{c}{600.}\\
NGC6822-HV-LBFLD1B (o6642) & 19 44 52.12  & -14 43 15.0 & \multicolumn{2}{c}{1320.} & \multicolumn{3}{c}{480.}\\
\enddata
\end{deluxetable}
}}
\normalsize


\renewcommand{\arraystretch}{.6}
\begin{deluxetable}{llllllll}
\tabletypesize{\tiny}
\tablewidth{0pt}
\tablecaption{ Magnitude limits and number of stars for different photometric error limits
\label{tbl:errs} }
\tablehead{
\\
\multicolumn{1}{c}{Error}  & \multicolumn{7}{c}{Magnitude limit (number of stars)} \\
\multicolumn{1}{c}{}  & \multicolumn{5}{c}{} & \multicolumn{2}{c}{}\\
\multicolumn{1}{c}{[mag]}  & \multicolumn{5}{c}{------- WFPC2 -------} & \multicolumn{2}{c}{------- STIS -------}\\
\multicolumn{1}{c}{}  & \multicolumn{5}{c}{} & \multicolumn{2}{c}{}\\
\multicolumn{1}{c}{}  & \multicolumn{1}{c}{F170W} & \multicolumn{1}{c}{F255W} & \multicolumn{1}{c}{F336W} & 
\multicolumn{1}{c}{F439W} & \multicolumn{1}{c}{F555W} & \multicolumn{1}{c}{FUV-MAMA}  & \multicolumn{1}{c}{NUV-MAMA} \\ 
}
\startdata
\\
$<$0.2	& 19.6 (105)    & 20.5 (100)	& 22.3 (536)	& 23.7 (1757)	& 24.0 (12423)& 22.0 (79) & 22.5 (80)\\
$<$0.1	& 18.6 (54)	& 19.5 (57)	& 21.4 (269)	& 22.8 (616)	& 23.3 (4946) & 21.5 (61) & 22.5 (80)\\
$<$0.05	& 17.8 (19)	& 18.5 (23)	& 20.3 (129)	& 21.8 (302)	& 22.3 (1386) & 20.5 (45) & 21.5 (67)\\

\enddata
\end{deluxetable}
\normalsize


{\rotate		
\renewcommand{\arraystretch}{.6}
{\footnotesize
\begin{deluxetable}{l@{\ \ }r@{\ \ }r@{\ \ }r@{\ \ }r@{\ \ }r@{\ \ }r@{\ \ }r@{\ \ }r@{\ \ }r@{\ \ }r@{\ \ }r@{\ \ }l}
\tabletypesize{\tiny}
\tablewidth{0pt}
\tablecaption{WFPC2 photometry for the hottest stars \tablenotemark{a} }.
\label{tbl:hot}
\tablehead{
\\
\multicolumn{1}{c}{Num.} & \multicolumn{1}{c}{RA (J2000)} & \multicolumn{1}{c}{Dec (J2000)}   & \multicolumn{1}{c}{\magv}   & \multicolumn{1}{c}{\magb} & \multicolumn{1}{c}{\magu}  & \multicolumn{1}{c}{\magn} & \multicolumn{1}{c}{\magf}  & \multicolumn{1}{c}{V}  & \multicolumn{1}{c}{B}  &\multicolumn{1}{c}{ U} & \multicolumn{1}{c}{\ebv}  & \multicolumn{1}{c}{\Teff [K]}\\

}
\startdata

     $BE\_WFPC2\_1$ & 19 44 37.145 & -14 39 49.48 &      20.94$\pm0.03$ &      21.15$\pm0.04$ &      19.55$\pm0.04$ &      19.00$\pm0.11$ &      19.09$\pm0.20$ & 20.93 & 21.15 & 20.22 &     0.32$\pm0.03$ &   $42700.^{+7500.}_{-4900.}$\\
     $BE\_WFPC2\_2$ & 19 45 03.565 & -14 46 14.96 &      21.11$\pm0.03$ &      21.12$\pm0.03$ &      19.98$\pm0.04$ &      19.93$\pm0.09$ &      19.75$\pm0.12$ & 21.11 & 21.12 & 20.48 &     0.27$\pm0.05$ &   $22600.^{+3000.}_{-1900.}$\\
     $BE\_WFPC2\_3$ & 19 45 06.802 & -14 47 04.43 &      20.39$\pm0.03$ &      20.40$\pm0.03$ &      19.14$\pm0.03$ &      18.93$\pm0.05$ &      18.70$\pm0.05$ & 20.39 & 20.40 & 19.69 &     0.27$\pm0.04$ &   $26200.^{+3000.}_{-1900.}$\\
     $BE\_WFPC2\_4$ & 19 45 06.989 & -14 47 01.86 &      21.32$\pm0.03$ &      21.35$\pm0.04$ &      20.13$\pm0.04$ &      19.99$\pm0.10$ &      19.81$\pm0.14$ & 21.32 & 21.35 & 20.66 &     0.30$\pm0.06$ &   $26800.^{+4000.}_{-2900.}$\\
     $BE\_WFPC2\_5$ & 19 45 08.502 & -14 47 09.98 &      20.42$\pm0.03$ &      20.44$\pm0.03$ &      19.01$\pm0.03$ &      18.70$\pm0.05$ &      18.25$\pm0.05$ & 20.42 & 20.44 & 19.62 &     0.29$\pm0.04$ &   $41200.^{+5000.}_{-7400.}$\\
     $BE\_WFPC2\_6$ & 19 45 08.604 & -14 46 26.46 &      21.05$\pm0.03$ &      21.12$\pm0.03$ &      20.02$\pm0.04$ &      20.13$\pm0.12$ &      19.80$\pm0.13$ & 21.05 & 21.12 & 20.50 &     0.31$\pm0.06$ &   $22700.^{+3000.}_{-1900.}$\\
     $BE\_WFPC2\_7$ & 19 45 09.247 & -14 45 51.89 &      20.24$\pm0.03$ &      20.21$\pm0.03$ &      18.72$\pm0.03$ &      18.39$\pm0.05$ &      18.03$\pm0.06$ & 20.24 & 20.21 & 19.35 &     0.25$\pm0.03$ &   $36900.^{+7500.}_{-2900.}$\\
     $BE\_WFPC2\_8$ & 19 45 09.408 & -14 47 10.14 &      21.49$\pm0.03$ &      21.52$\pm0.04$ &      20.27$\pm0.05$ &      20.15$\pm0.12$ &      19.69$\pm0.11$ & 21.49 & 21.52 & 20.81 &     0.28$\pm0.07$ &  $29700.^{+10500.}_{-3900.}$\\
     $BE\_WFPC2\_9$ & 19 45 10.140 & -14 44 12.39 &      21.71$\pm0.04$ &      21.76$\pm0.05$ &      20.49$\pm0.06$ &      20.24$\pm0.20$ &      19.55$\pm0.19$ & 21.71 & 21.76 & 21.04 &     0.27$\pm0.08$ &  $33400.^{+14000.}_{-6900.}$\\
    $BE\_WFPC2\_10$ & 19 45 11.411 & -14 44 24.89 &      20.92$\pm0.03$ &      20.92$\pm0.03$ &      19.63$\pm0.04$ &      19.07$\pm0.08$ &      18.82$\pm0.10$ & 20.92 & 20.92 & 20.19 &     0.16$\pm0.05$ &   $24900.^{+4000.}_{-1900.}$\\
    $BE\_WFPC2\_11$ & 19 45 11.836 & -14 44 22.31 &      20.46$\pm0.03$ &      20.42$\pm0.03$ &      19.13$\pm0.03$ &      18.78$\pm0.06$ &      18.42$\pm0.07$ & 20.46 & 20.42 & 19.69 &     0.18$\pm0.05$ &   $24900.^{+2000.}_{-2900.}$\\
    $BE\_WFPC2\_12$ & 19 45 12.206 & -14 46 59.01 &      20.72$\pm0.03$ &      20.81$\pm0.03$ &      19.37$\pm0.03$ &      19.32$\pm0.09$ &      18.90$\pm0.10$ & 20.71 & 20.81 & 19.98 &     0.37$\pm0.03$ &   $42400.^{+5000.}_{-7400.}$\\
    $BE\_WFPC2\_13$ & 19 45 12.229 & -14 46 58.89 &      20.77$\pm0.03$ &      20.78$\pm0.03$ &      19.48$\pm0.03$ &      19.51$\pm0.07$ &      19.16$\pm0.10$ & 20.77 & 20.78 & 20.04 &     0.36$\pm0.05$ &   $34300.^{+9500.}_{-5900.}$\\
    $BE\_WFPC2\_14$ & 19 45 12.250 & -14 44 47.56 &      19.93$\pm0.02$ &      19.95$\pm0.03$ &      18.64$\pm0.03$ &      18.55$\pm0.05$ &      18.10$\pm0.06$ & 19.93 & 19.95 & 19.20 &     0.36$\pm0.05$ &   $39400.^{+7500.}_{-9900.}$\\
    $BE\_WFPC2\_15$ & 19 45 12.657 & -14 45 00.28 &      19.42$\pm0.02$ &      19.42$\pm0.03$ &      18.14$\pm0.03$ &      18.02$\pm0.04$ &      17.80$\pm0.05$ & 19.42 & 19.42 & 18.69 &     0.32$\pm0.04$ &   $28900.^{+4000.}_{-2900.}$\\
    $BE\_WFPC2\_16$ & 19 45 12.768 & -14 45 20.39 &      20.92$\pm0.03$ &      20.97$\pm0.03$ &      19.63$\pm0.03$ &      19.39$\pm0.09$ &      19.06$\pm0.12$ & 20.92 & 20.97 & 20.21 &     0.31$\pm0.05$ &  $35000.^{+12000.}_{-4900.}$\\
    $BE\_WFPC2\_17$ & 19 45 12.827 & -14 45 17.53 &      19.46$\pm0.02$ &      19.51$\pm0.03$ &      18.23$\pm0.03$ &      17.98$\pm0.04$ &      18.08$\pm0.06$ & 19.46 & 19.51 & 18.78 &     0.32$\pm0.04$ &    $27100.^{+4000.}_{-900.}$\\
    $BE\_WFPC2\_18$ & 19 45 12.972 & -14 43 00.83 &      20.36$\pm0.03$ &      20.30$\pm0.03$ &      18.89$\pm0.03$ &      18.61$\pm0.06$ &      17.90$\pm0.05$ & 20.36 & 20.30 & 19.49 &     0.27$\pm0.04$ &   $45100.^{+2500.}_{-7400.}$\\
    $BE\_WFPC2\_19$ & 19 45 13.001 & -14 43 00.80 &      20.24$\pm0.03$ &      20.24$\pm0.03$ &      18.84$\pm0.03$ &      18.66$\pm0.05$ &      18.19$\pm0.06$ & 20.24 & 20.24 & 19.44 &     0.32$\pm0.04$ &   $41000.^{+5000.}_{-8400.}$\\
    $BE\_WFPC2\_20$ & 19 45 13.310 & -14 45 32.98 &      21.49$\pm0.03$ &      21.49$\pm0.04$ &      20.12$\pm0.04$ &      19.68$\pm0.11$ &      19.57$\pm0.19$ & 21.49 & 21.49 & 20.71 &     0.24$\pm0.06$ &  $32200.^{+10000.}_{-3900.}$\\
    $BE\_WFPC2\_21$ & 19 45 13.319 & -14 44 51.42 &      20.08$\pm0.03$ &      20.08$\pm0.03$ &      18.61$\pm0.03$ &      18.27$\pm0.04$ &      17.85$\pm0.05$ & 20.08 & 20.08 & 19.23 &     0.27$\pm0.03$ &   $39200.^{+5000.}_{-6900.}$\\
    $BE\_WFPC2\_22$ & 19 45 13.356 & -14 45 25.61 &      20.08$\pm0.03$ &      20.04$\pm0.03$ &      18.53$\pm0.03$ &      18.24$\pm0.04$ &      17.75$\pm0.05$ & 20.08 & 20.04 & 19.17 &     0.27$\pm0.03$ &   $41500.^{+5000.}_{-7400.}$\\
    $BE\_WFPC2\_23$ & 19 45 13.435 & -14 44 38.02 &      20.24$\pm0.03$ &      20.23$\pm0.03$ &      18.84$\pm0.03$ &      18.69$\pm0.06$ &      18.11$\pm0.07$ & 20.24 & 20.23 & 19.43 &     0.33$\pm0.04$ &   $43000.^{+5000.}_{-7400.}$\\
    $BE\_WFPC2\_24$ & 19 45 13.459 & -14 44 38.39 &      19.22$\pm0.02$ &      19.35$\pm0.02$ &      17.88$\pm0.02$ &      17.77$\pm0.04$ &      17.47$\pm0.04$ & 19.21 & 19.35 & 18.50 &     0.39$\pm0.03$ &   $42000.^{+5000.}_{-4900.}$\\
    $BE\_WFPC2\_25$ & 19 45 13.551 & -14 45 11.10 &      20.03$\pm0.03$ &      20.24$\pm0.03$ &      18.77$\pm0.03$ &      18.62$\pm0.05$ &      18.48$\pm0.08$ & 20.02 & 20.24 & 19.39 &     0.41$\pm0.03$ &   $41600.^{+5000.}_{-7400.}$\\
    $BE\_WFPC2\_26$ & 19 45 13.590 & -14 45 12.82 &      18.98$\pm0.02$ &      19.07$\pm0.02$ &      17.58$\pm0.02$ &      17.42$\pm0.03$ &      16.84$\pm0.03$ & 18.97 & 19.07 & 18.21 &     0.35$\pm0.01$ &      $48300.^{+0.}_{-4900.}$\\
    $BE\_WFPC2\_27$ & 19 45 14.388 & -14 44 52.07 &      20.92$\pm0.03$ &      20.94$\pm0.03$ &      19.60$\pm0.04$ &      19.27$\pm0.08$ &      18.89$\pm0.10$ & 20.92 & 20.94 & 20.18 &     0.26$\pm0.06$ &  $32300.^{+10000.}_{-3900.}$\\
    $BE\_WFPC2\_28$ & 19 45 14.529 & -14 45 47.99 &      20.38$\pm0.03$ &      20.38$\pm0.03$ &      19.06$\pm0.03$ &      18.80$\pm0.06$ &      18.45$\pm0.08$ & 20.38 & 20.38 & 19.63 &     0.26$\pm0.05$ &   $29500.^{+6000.}_{-2900.}$\\
    $BE\_WFPC2\_29$ & 19 45 14.797 & -14 44 53.35 &      21.04$\pm0.03$ &      21.04$\pm0.03$ &      19.66$\pm0.04$ &      19.49$\pm0.10$ &      19.17$\pm0.14$ & 21.04 & 21.04 & 20.25 &     0.30$\pm0.05$ &  $35800.^{+10000.}_{-5900.}$\\
    $BE\_WFPC2\_30$ & 19 45 14.889 & -14 45 17.63 &      20.65$\pm0.03$ &      20.68$\pm0.03$ &      19.32$\pm0.03$ &      18.85$\pm0.07$ &      18.55$\pm0.09$ & 20.65 & 20.68 & 19.90 &     0.22$\pm0.05$ &   $29900.^{+6000.}_{-2900.}$\\
    $BE\_WFPC2\_31$ & 19 45 14.978 & -14 45 08.35 &      21.11$\pm0.03$ &      21.17$\pm0.04$ &      19.98$\pm0.04$ &      19.59$\pm0.11$ &      19.33$\pm0.19$ & 21.11 & 21.17 & 20.50 &     0.24$\pm0.06$ &   $24200.^{+3000.}_{-2900.}$\\
    $BE\_WFPC2\_32$ & 19 45 15.103 & -14 45 04.31 &      20.10$\pm0.03$ &      20.10$\pm0.03$ &      18.82$\pm0.03$ &      18.50$\pm0.05$ &      18.33$\pm0.07$ & 20.10 & 20.10 & 19.37 &     0.24$\pm0.04$ &   $25700.^{+3000.}_{-1900.}$\\
    $BE\_WFPC2\_33$ & 19 45 15.125 & -14 45 11.46 &      18.59$\pm0.02$ &      18.69$\pm0.02$ &      17.47$\pm0.02$ &      17.37$\pm0.03$ &      17.50$\pm0.05$ & 18.58 & 18.69 & 18.00 &     0.39$\pm0.04$ &    $26900.^{+3000.}_{-900.}$\\
    $BE\_WFPC2\_34$ & 19 45 15.193 & -14 45 11.53 &      20.57$\pm0.03$ &      20.62$\pm0.03$ &      19.10$\pm0.03$ &      18.87$\pm0.06$ &      18.21$\pm0.07$ & 20.57 & 20.62 & 19.74 &     0.30$\pm0.02$ &   $45800.^{+2500.}_{-4900.}$\\
    $BE\_WFPC2\_35$ & 19 45 15.367 & -14 45 33.47 &      21.11$\pm0.03$ &      21.17$\pm0.04$ &      19.67$\pm0.04$ &      19.55$\pm0.10$ &      19.23$\pm0.17$ & 21.11 & 21.17 & 20.30 &     0.33$\pm0.03$ &   $41600.^{+5000.}_{-7400.}$\\
    $BE\_WFPC2\_36$ & 19 45 15.454 & -14 45 33.65 &      21.22$\pm0.03$ &      21.39$\pm0.04$ &      19.95$\pm0.04$ &      19.61$\pm0.12$ &      19.25$\pm0.18$ & 21.21 & 21.39 & 20.56 &     0.36$\pm0.04$ &   $41700.^{+5000.}_{-7400.}$\\
    $BE\_WFPC2\_37$ & 19 45 15.521 & -14 45 21.51 &      20.58$\pm0.03$ &      20.52$\pm0.03$ &      19.03$\pm0.03$ &      18.75$\pm0.06$ &      18.17$\pm0.06$ & 20.58 & 20.52 & 19.66 &     0.25$\pm0.04$ &   $41900.^{+5000.}_{-7400.}$\\
    $BE\_WFPC2\_38$ & 19 45 15.618 & -14 45 23.69 &      18.66$\pm0.02$ &      18.63$\pm0.02$ &      17.10$\pm0.02$ &      16.85$\pm0.03$ &      16.41$\pm0.03$ & 18.66 & 18.63 & 17.75 &     0.28$\pm0.03$ &   $40000.^{+5000.}_{-4900.}$\\
    $BE\_WFPC2\_39$ & 19 45 15.638 & -14 45 20.66 &      20.23$\pm0.03$ &      20.23$\pm0.03$ &      18.76$\pm0.03$ &      18.46$\pm0.05$ &      18.10$\pm0.06$ & 20.23 & 20.23 & 19.38 &     0.28$\pm0.03$ &   $38000.^{+7500.}_{-4400.}$\\
    $BE\_WFPC2\_40$ & 19 45 15.685 & -14 45 06.49 &      20.33$\pm0.03$ &      20.29$\pm0.03$ &      18.84$\pm0.03$ &      18.54$\pm0.05$ &      18.00$\pm0.06$ & 20.33 & 20.29 & 19.46 &     0.27$\pm0.04$ &   $41800.^{+5000.}_{-7400.}$\\
    $BE\_WFPC2\_41$ & 19 45 15.737 & -14 45 32.75 &      20.60$\pm0.03$ &      20.65$\pm0.03$ &      19.29$\pm0.03$ &      19.02$\pm0.07$ &      18.41$\pm0.08$ & 20.60 & 20.65 & 19.87 &     0.31$\pm0.05$ &  $42300.^{+5000.}_{-11900.}$\\
    $BE\_WFPC2\_42$ & 19 45 15.908 & -14 45 33.89 &      20.91$\pm0.03$ &      20.87$\pm0.03$ &      19.43$\pm0.03$ &      19.00$\pm0.07$ &      18.74$\pm0.10$ & 20.91 & 20.87 & 20.04 &     0.22$\pm0.05$ &   $33400.^{+8000.}_{-3900.}$\\
    $BE\_WFPC2\_43$ & 19 45 15.922 & -14 44 58.13 &      19.54$\pm0.02$ &      19.55$\pm0.03$ &      18.10$\pm0.03$ &      17.85$\pm0.04$ &      17.56$\pm0.04$ & 19.54 & 19.55 & 18.72 &     0.30$\pm0.02$ &   $35200.^{+3500.}_{-1900.}$\\
    $BE\_WFPC2\_44$ & 19 45 16.115 & -14 45 38.47 &      20.58$\pm0.03$ &      20.61$\pm0.03$ &      19.08$\pm0.03$ &      18.59$\pm0.06$ &      17.98$\pm0.06$ & 20.58 & 20.61 & 19.73 &     0.23$\pm0.03$ &   $44800.^{+5000.}_{-4900.}$\\
    $BE\_WFPC2\_45$ & 19 45 16.366 & -14 45 14.88 &      18.49$\pm0.02$ &      18.53$\pm0.02$ &      17.12$\pm0.02$ &      17.07$\pm0.03$ &      16.89$\pm0.03$ & 18.49 & 18.53 & 17.72 &     0.38$\pm0.01$ &    $34300.^{+1000.}_{-900.}$\\
    $BE\_WFPC2\_46$ & 19 45 16.832 & -14 45 07.13 &      20.74$\pm0.03$ &      20.78$\pm0.03$ &      19.32$\pm0.03$ &      18.98$\pm0.07$ &      18.50$\pm0.10$ & 20.74 & 20.78 & 19.94 &     0.29$\pm0.04$ &   $41500.^{+5000.}_{-7400.}$\\
    $BE\_WFPC2\_47$ & 19 45 16.941 & -14 45 22.50 &      21.27$\pm0.03$ &      21.30$\pm0.04$ &      19.98$\pm0.04$ &      19.42$\pm0.12$ &      19.11$\pm0.14$ & 21.27 & 21.30 & 20.55 &     0.21$\pm0.06$ &  $29400.^{+10500.}_{-2900.}$\\
\enddata
\tablenotetext{a}{The stars with measurements in all five WFPC2 bands are given in 
the printed version, the photometry for the entire sample is available in the electronic version.
The 
first star is in the ``Group 2'' fields, all the others are in the ``Group 1'' fields. 
Magnitudes are in the Vega-mag system.} 
\end{deluxetable}
}}
\normalsize

\renewcommand{\arraystretch}{.6}
{\footnotesize
\begin{deluxetable}{l@{\ \ }r@{\ \ }r@{\ \ }r@{\ \ }r@{\ \ }r@{\ \ }l@{\ \ }l}
\tabletypesize{\tiny}
\tablewidth{0pt}
\tablecaption{STIS photometry. 
\label{tbl:hotstis}}
\tablehead{
\\
\multicolumn{1}{c}{Name} & \multicolumn{1}{c}{RA (J2000)} & \multicolumn{1}{c}{Dec (J2000)} & \multicolumn{1}{c}{NUV} & \multicolumn{1}{c}{FUV} & \multicolumn{1}{c}{\ebv}  & \multicolumn{1}{c}{\Teff [K]} & \multicolumn{1}{c}{Identification\tablenotemark{a}}\\

}
\startdata
     $BE\_STIS\_1$ & 19 44 51.197 & -14 43 08.67 &      22.03$\pm0.06$ &      21.29$\pm0.12$ &              ---- &                         ---- &                ----\\
      $BE\_STIS\_2$ & 19 44 51.234 & -14 43 09.26 &      19.96$\pm0.02$ &      19.14$\pm0.02$ &     0.35$\pm0.05$ &   $26200.^{+2800.}_{-2200.}$ &           LB-f1-540\\
      $BE\_STIS\_3$ & 19 44 51.308 & -14 43 12.65 &      18.49$\pm0.01$ &      17.79$\pm0.01$ &     0.41$\pm0.01$ &   $38500.^{+1000.}_{-1000.}$ &           LB-f1-536\\
      $BE\_STIS\_4$ & 19 44 51.343 & -14 43 19.69 &      17.90$\pm0.01$ &      17.19$\pm0.01$ &     0.40$\pm0.01$ &    $37300.^{+2700.}_{-200.}$ &           LB-f1-525\\
      $BE\_STIS\_5$ & 19 44 51.351 & -14 43 06.87 &      22.07$\pm0.06$ &      21.48$\pm0.13$ &     0.43$\pm0.22$ &  $23900.^{+11100.}_{-8900.}$ &           LB-f1-555\\
      $BE\_STIS\_6$ & 19 44 51.380 & -14 43 08.42 &      21.63$\pm0.05$ &      21.10$\pm0.10$ &     0.34$\pm0.13$ &   $18900.^{+5100.}_{-2900.}$ &           LB-f1-549\\
      $BE\_STIS\_7$ & 19 44 51.387 & -14 43 07.07 &      21.83$\pm0.05$ &      21.40$\pm0.12$ &     0.20$\pm0.15$ &   $15000.^{+3000.}_{-2000.}$ &           LB-f1-558\\
      $BE\_STIS\_8$ & 19 44 51.388 & -14 43 18.82 &      21.84$\pm0.05$ &      21.19$\pm0.09$ &     0.11$\pm0.17$ &   $14000.^{+5000.}_{-2000.}$ &           LB-f1-527\\
      $BE\_STIS\_9$ & 19 44 51.550 & -14 43 19.79 &      19.16$\pm0.01$ &      18.31$\pm0.01$ &     0.37$\pm0.01$ &   $35100.^{+4900.}_{-1100.}$ &           LB-f1-531\\
     $BE\_STIS\_10$ & 19 44 51.630 & -14 43 06.44 &      22.08$\pm0.06$ &      21.43$\pm0.11$ &     0.32$\pm0.21$ &   $19100.^{+5900.}_{-5100.}$ &           LB-f1-578\\
     $BE\_STIS\_11$ & 19 44 51.667 & -14 43 19.51 &      19.02$\pm0.01$ &      18.24$\pm0.01$ &     0.40$\pm0.01$ &   $36600.^{+3400.}_{-2600.}$ &           LB-f1-535\\
     $BE\_STIS\_12$ & 19 44 51.667 & -14 43 13.09 &      19.71$\pm0.02$ &      18.96$\pm0.02$ &     0.41$\pm0.04$ &   $27200.^{+2800.}_{-3200.}$ &           LB-f1-552\\
     $BE\_STIS\_13$ & 19 44 51.681 & -14 43 13.61 &      20.54$\pm0.03$ &      19.73$\pm0.03$ &     0.37$\pm0.08$ &   $28900.^{+8600.}_{-5900.}$ &           LB-f1-550\\
     $BE\_STIS\_14$ & 19 44 51.696 & -14 43 10.88 &      20.63$\pm0.03$ &      19.99$\pm0.04$ &     0.31$\pm0.09$ &   $21900.^{+4100.}_{-2900.}$ &           LB-f1-566\\
     $BE\_STIS\_15$ & 19 44 51.725 & -14 43 18.91 &      17.10$\pm0.01$ &      16.39$\pm0.01$ &     0.41$\pm0.01$ &    $37200.^{+2800.}_{-300.}$ &           LB-f1-539\\
     $BE\_STIS\_16$ & 19 44 51.740 & -14 43 16.22 &      18.90$\pm0.01$ &      18.30$\pm0.01$ &     0.52$\pm0.01$ &   $39500.^{+3000.}_{-2000.}$ &           LB-f1-546\\
     $BE\_STIS\_17$ & 19 44 51.747 & -14 43 15.02 &      21.02$\pm0.03$ &      20.31$\pm0.05$ &     0.29$\pm0.13$ &   $23000.^{+9000.}_{-5000.}$ &           LB-f1-548\\
     $BE\_STIS\_18$ & 19 44 51.747 & -14 43 13.87 &      21.08$\pm0.03$ &      20.68$\pm0.05$ &     0.07$\pm0.02$ &   $14000.^{+1000.}_{-1000.}$ &           LB-f1-557\\
     $BE\_STIS\_19$ & 19 44 51.799 & -14 43 10.62 &      21.77$\pm0.05$ &      21.24$\pm0.10$ &     0.30$\pm0.15$ &   $18100.^{+3900.}_{-3100.}$ &           LB-f1-573\\
     $BE\_STIS\_20$ & 19 44 51.806 & -14 43 17.16 &      21.24$\pm0.04$ &      20.59$\pm0.06$ &     0.12$\pm0.11$ &   $15100.^{+2900.}_{-2100.}$ &           LB-f1-547\\
     $BE\_STIS\_21$ & 19 44 51.813 & -14 43 18.25 &      21.45$\pm0.04$ &      20.73$\pm0.07$ &     0.06$\pm0.08$ &   $15000.^{+3000.}_{-1000.}$ &           LB-f1-544\\
     $BE\_STIS\_22$ & 19 44 51.821 & -14 43 07.57 &      16.82$\pm0.00$ &      16.52$\pm0.01$ &     0.51$\pm0.01$ &     $38300.^{+800.}_{-800.}$ &           LB-f1-584\\
     $BE\_STIS\_23$ & 19 44 51.864 & -14 43 15.50 &      19.57$\pm0.02$ &      18.82$\pm0.02$ &     0.26$\pm0.05$ &   $23200.^{+2800.}_{-3200.}$ &           LB-f1-559\\
     $BE\_STIS\_24$ & 19 44 51.872 & -14 43 12.44 &      20.68$\pm0.03$ &      20.07$\pm0.04$ &     0.39$\pm0.10$ &   $25800.^{+8200.}_{-4800.}$ &           LB-f1-571\\
     $BE\_STIS\_25$ & 19 44 51.894 & -14 43 20.21 &      20.84$\pm0.03$ &      20.08$\pm0.04$ &     0.48$\pm0.03$ &  $45000.^{+5000.}_{-12000.}$ &           LB-f1-541\\
     $BE\_STIS\_26$ & 19 44 51.894 & -14 43 19.68 &      22.20$\pm0.06$ &      21.46$\pm0.10$ &     0.46$\pm0.06$ & $38600.^{+11400.}_{-11600.}$ &           LB-f1-543\\
     $BE\_STIS\_27$ & 19 44 51.938 & -14 43 19.86 &      19.79$\pm0.02$ &      19.05$\pm0.02$ &     0.42$\pm0.04$ &  $35000.^{+15000.}_{-4000.}$ &           LB-f1-545\\
     $BE\_STIS\_28$ & 19 44 51.989 & -14 43 14.81 &      17.88$\pm0.01$ &      17.23$\pm0.01$ &     0.41$\pm0.01$ &   $38700.^{+1300.}_{-1200.}$ &           LB-f1-569\\
     $BE\_STIS\_29$ & 19 44 52.011 & -14 43 05.20 &      17.24$\pm0.01$ &      16.73$\pm0.01$ &     0.43$\pm0.01$ &     $38400.^{+900.}_{-900.}$ &           LB-f1-597\\
     $BE\_STIS\_30$ & 19 44 52.018 & -14 43 17.91 &      20.79$\pm0.03$ &      20.13$\pm0.04$ &     0.18$\pm0.09$ &   $17000.^{+3000.}_{-2000.}$ &           LB-f1-561\\
     $BE\_STIS\_31$ & 19 44 52.055 & -14 43 17.83 &      21.26$\pm0.04$ &      20.63$\pm0.06$ &     0.49$\pm0.07$ &   $28000.^{+5000.}_{-7000.}$ &           LB-f1-563\\
     $BE\_STIS\_32$ & 19 44 52.055 & -14 43 14.87 &      17.00$\pm0.01$ &      16.34$\pm0.01$ &     0.41$\pm0.01$ &     $38300.^{+800.}_{-800.}$ &           LB-f1-575\\
     $BE\_STIS\_33$ & 19 44 52.062 & -14 43 14.20 &      21.55$\pm0.05$ &      21.10$\pm0.11$ &     0.24$\pm0.15$ &   $15100.^{+2900.}_{-2100.}$ &           LB-f1-579\\
     $BE\_STIS\_34$ & 19 44 52.062 & -14 43 02.12 &      17.99$\pm0.01$ &      17.49$\pm0.01$ &              ---- &                         ---- &                ----\\
     $BE\_STIS\_35$ & 19 44 52.070 & -14 43 16.57 &      21.25$\pm0.04$ &      20.53$\pm0.05$ &     0.47$\pm0.05$ & $38200.^{+11800.}_{-16200.}$ &           LB-f1-567\\
     $BE\_STIS\_36$ & 19 44 52.091 & -14 43 02.50 &      19.52$\pm0.02$ &      19.23$\pm0.03$ &     0.30$\pm0.01$ &   $18000.^{+1000.}_{-1000.}$ &           LB-f1-609\\
     $BE\_STIS\_37$ & 19 44 52.128 & -14 43 08.46 &      20.16$\pm0.02$ &      19.45$\pm0.03$ &     0.43$\pm0.05$ &  $34200.^{+13300.}_{-4200.}$ &           LB-f1-593\\
     $BE\_STIS\_38$ & 19 44 52.187 & -14 43 09.37 &      17.49$\pm0.01$ &      16.79$\pm0.01$ &     0.41$\pm0.01$ &     $40100.^{+100.}_{-100.}$ &           LB-f1-594\\
     $BE\_STIS\_39$ & 19 44 52.238 & -14 43 18.27 &      20.30$\pm0.02$ &      19.54$\pm0.03$ &     0.24$\pm0.06$ &   $19900.^{+4100.}_{-1900.}$ &           LB-f1-574\\
     $BE\_STIS\_40$ & 19 44 52.253 & -14 43 12.37 &      22.39$\pm0.06$ &      21.62$\pm0.14$ &     0.14$\pm0.05$ &    $15600.^{+2300.}_{-700.}$ &           LB-f1-591\\
     $BE\_STIS\_41$ & 19 44 52.282 & -14 43 14.85 &      20.48$\pm0.03$ &      19.75$\pm0.03$ &     0.43$\pm0.03$ &   $38100.^{+9400.}_{-5100.}$ &           LB-f1-586\\
     $BE\_STIS\_42$ & 19 44 52.304 & -14 43 01.53 &      21.31$\pm0.04$ &      21.32$\pm0.11$ &     0.27: &   $16100.:$ &           LB-f1-621\\
     $BE\_STIS\_43$ & 19 44 52.319 & -14 43 16.19 &      20.91$\pm0.03$ &      20.14$\pm0.05$ &     0.45$\pm0.08$ &  $31800.^{+18200.}_{-8800.}$ &           LB-f1-585\\
     $BE\_STIS\_44$ & 19 44 52.348 & -14 43 15.88 &      17.82$\pm0.01$ &      17.06$\pm0.01$ &     0.40$\pm0.01$ &     $38300.^{+800.}_{-800.}$ &           LB-f1-587\\
     $BE\_STIS\_45$ & 19 44 52.370 & -14 43 10.28 &      19.03$\pm0.01$ &      18.28$\pm0.01$ &     0.29$\pm0.01$ &   $23900.^{+3100.}_{-1900.}$ &           LB-f1-603\\
     $BE\_STIS\_46$ & 19 44 52.384 & -14 43 03.97 &      20.45$\pm0.02$ &      20.54$\pm0.06$ &     0.53$\pm0.08$ &   $24100.^{+3900.}_{-3100.}$ &           LB-f1-618\\
     $BE\_STIS\_47$ & 19 44 52.399 & -14 43 12.80 &      22.27$\pm0.06$ &      21.43$\pm0.11$ &     0.46$\pm0.17$ & $26300.^{+16200.}_{-10300.}$ &           LB-f1-596\\
     $BE\_STIS\_48$ & 19 44 52.399 & -14 43 12.24 &      20.33$\pm0.02$ &      19.64$\pm0.03$ &     0.28$\pm0.09$ &   $20900.^{+4100.}_{-2900.}$ &           LB-f1-598\\
     $BE\_STIS\_49$ & 19 44 52.406 & -14 43 15.61 &      20.00$\pm0.02$ &      19.26$\pm0.03$ &     0.43$\pm0.05$ &  $35100.^{+7400.}_{-10100.}$ &           LB-f1-589\\
     $BE\_STIS\_50$ & 19 44 52.414 & -14 43 11.82 &      22.12$\pm0.05$ &      21.47$\pm0.12$ &     :$$ &   :$$ &           LB-f1-601\\
     $BE\_STIS\_51$ & 19 44 52.487 & -14 43 14.25 &      21.85$\pm0.05$ &      21.33$\pm0.11$ &              ---- &                         ---- &                ----\\
     $BE\_STIS\_52$ & 19 44 52.502 & -14 43 13.96 &      21.30$\pm0.04$ &      20.60$\pm0.06$ &     0.17$\pm0.12$ &   $17000.^{+3000.}_{-2000.}$ &           LB-f1-599\\
     $BE\_STIS\_53$ & 19 44 52.502 & -14 43 13.38 &      20.25$\pm0.02$ &      19.60$\pm0.03$ &     0.25$\pm0.06$ &   $20100.^{+2900.}_{-3100.}$ &           LB-f1-602\\
     $BE\_STIS\_54$ & 19 44 52.516 & -14 43 12.47 &      21.84$\pm0.05$ &      21.29$\pm0.10$ &     0.55$\pm0.12$ & $31200.^{+18800.}_{-11200.}$ &           LB-f1-604\\
     $BE\_STIS\_55$ & 19 44 52.648 & -14 43 12.07 &      21.83$\pm0.05$ &      21.48$\pm0.11$ &              ---- &                         ---- &                ----\\
     $BE\_STIS\_56$ & 19 44 52.670 & -14 43 11.65 &      18.51$\pm0.01$ &      17.76$\pm0.01$ &     0.40$\pm0.01$ &     $37900.^{+400.}_{-400.}$ &           LB-f1-610\\
     $BE\_STIS\_57$ & 19 44 52.721 & -14 43 21.44 &      22.12$\pm0.05$ &      21.46$\pm0.11$ &              ---- &                         ---- &                ----\\
     $BE\_STIS\_58$ & 19 44 52.765 & -14 43 12.02 &      18.24$\pm0.01$ &      18.34$\pm0.01$ &     0.63$:$ &    $39500.:$ &           LB-f1-614\\
     $BE\_STIS\_59$ & 19 44 52.824 & -14 43 20.67 &      21.88$\pm0.05$ &      21.24$\pm0.10$ &     0.40$\pm0.15$ &   $21800.^{+6200.}_{-4800.}$ &           LB-f1-595\\

\enddata
\tablenotetext{a}{The last column provides the cross-identification with the WFPC2 photometric study
 by Bianchi et al. (2001). Magnitudes are in the Vega-mag system. }
\end{deluxetable}
}
\normalsize

\renewcommand{\arraystretch}{.6}
{\footnotesize
\begin{deluxetable}{r@{\ \ }r@{\ \ }r@{\ \ }r@{\ \ }r@{\ \ }r@{\ \ }r@{\ \ }r}
\tabletypesize{\tiny}
\tablenum{4}
\tablewidth{0pt}
\tablecaption{ STIS photometry - continued.
\label{tbl:hotstis2}}
\tablehead{
\\
\multicolumn{1}{c}{Name} & \multicolumn{1}{c}{RA (J2000)} & \multicolumn{1}{c}{Dec (J2000)} & \multicolumn{1}{c}{NUV} & \multicolumn{1}{c}{FUV} & \multicolumn{1}{c}{\ebv}  & \multicolumn{1}{c}{\Teff [K]} & \multicolumn{1}{c}{ID}\\

}
\startdata

     $BE\_STIS\_60$ & 19 44 52.831 & -14 43 12.89 &      20.84$\pm0.03$ &      20.23$\pm0.05$ &     0.68$:$ &     $50000.:$ &           LB-f1-616\\
     $BE\_STIS\_61$ & 19 44 52.897 & -14 43 11.38 &      21.14$\pm0.04$ &      21.31$\pm0.13$ &     0.45$\pm0.31$ &   $18100.^{+5900.}_{-5100.}$ &           LB-f1-626\\
     $BE\_STIS\_62$ & 19 44 52.904 & -14 43 12.20 &      17.08$\pm0.00$ &      16.72$\pm0.01$ &     0.50$\pm0.01$ &    $37900.^{+2100.}_{-400.}$ &           LB-f1-622\\
     $BE\_STIS\_63$ & 19 44 52.904 & -14 43 09.81 &      22.06$\pm0.05$ &      21.71$\pm0.14$ &     0.57$\pm0.25$ &  $22100.^{+10900.}_{-9100.}$ &           LB-f1-635\\
     $BE\_STIS\_64$ & 19 44 52.926 & -14 43 13.35 &      22.07$\pm0.07$ &      21.75$\pm0.16$ &     :$$ &   $:$ &           LB-f1-619\\
     $BE\_STIS\_65$ & 19 44 52.934 & -14 43 11.80 &      20.06$\pm0.02$ &      19.91$\pm0.05$ &     0.47$\pm0.15$ &   $22100.^{+6900.}_{-4100.}$ &           LB-f1-628\\
     $BE\_STIS\_66$ & 19 44 52.948 & -14 43 12.38 &            22.07$\pm0.09$ &      21.99$\pm0.28$ &     ---- &  ---- &        ----\\
     $BE\_STIS\_67$ & 19 44 52.956 & -14 43 11.03 &      20.45$\pm0.03$ &      20.29$\pm0.05$ &     0.65$\pm0.06$ & $38200.^{+11800.}_{-18200.}$ &           LB-f1-632\\
     $BE\_STIS\_68$ & 19 44 52.963 & -14 43 12.35 &      19.56$\pm0.02$ &      19.04$\pm0.02$ &     0.58$\pm0.02$ &   $38300.^{+4200.}_{-3300.}$ &           LB-f1-627\\
     $BE\_STIS\_69$ & 19 44 52.963 & -14 43 11.70 &      20.07$\pm0.02$ &      19.62$\pm0.04$ &     0.61$\pm0.02$ &   $40200.^{+4800.}_{-2700.}$ &           LB-f1-630\\
     $BE\_STIS\_70$ & 19 44 52.978 & -14 43 13.41 &      21.02$\pm0.04$ &      20.52$\pm0.06$ &              ---- &                         ---- &                ----\\
     $BE\_STIS\_71$ & 19 44 52.978 & -14 43 11.67 &      19.79$\pm0.02$ &      19.37$\pm0.03$ &              ---- &                         ---- &                ----\\
     $BE\_STIS\_72$ & 19 44 52.985 & -14 43 11.16 &      19.65$\pm0.02$ &      19.55$\pm0.03$ &     0.62$\pm0.03$ &   $40800.^{+9200.}_{-6800.}$ &           LB-f1-636\\
     $BE\_STIS\_73$ & 19 44 53.000 & -14 43 12.75 &      18.23$\pm0.01$ &      17.62$\pm0.01$ &     0.47$\pm0.02$ &   $37000.^{+5500.}_{-3000.}$ &           LB-f1-629\\
     $BE\_STIS\_74$ & 19 44 53.022 & -14 43 12.17 &      20.19$\pm0.02$ &      19.92$\pm0.04$ &     0.73$\pm0.06$ &  $45100.^{+4900.}_{-16100.}$ &           LB-f1-634\\
     $BE\_STIS\_75$ & 19 44 53.029 & -14 43 11.34 &      20.43$\pm0.03$ &      20.32$\pm0.05$ &     0.70$\pm0.04$ &  $43100.^{+6900.}_{-14100.}$ &           LB-f1-640\\
     $BE\_STIS\_76$ & 19 44 53.044 & -14 43 12.07 &      21.11$\pm0.04$ &      21.23$\pm0.12$ &     0.76$\pm0.03$ &  $40000.^{+10000.}_{-6000.}$ &           LB-f1-637\\
     $BE\_STIS\_77$ & 19 44 53.073 & -14 43 11.26 &      19.86$\pm0.02$ &      19.73$\pm0.03$ &     0.64$\pm0.04$ &  $37000.^{+13000.}_{-5000.}$ &           LB-f1-644\\
     $BE\_STIS\_78$ & 19 44 53.175 & -14 43 12.92 &      20.16$\pm0.02$ &      19.80$\pm0.04$ &     0.53$\pm0.02$ &   $35100.^{+7400.}_{-2100.}$ &           LB-f1-646\\
     $BE\_STIS\_79$ & 19 44 53.183 & -14 43 12.48 &      21.16$\pm0.03$ &      20.74$\pm0.07$ &     0.43$\pm0.10$ &    $22000.^{+10000.}_{-10000.}$ &           LB-f1-648\\
     $BE\_STIS\_80$ & 19 44 53.219 & -14 43 12.29 &      21.52$\pm0.04$ &      21.12$\pm0.11$ &     0.44$\pm0.16$ &   $21200.^{+6800.}_{-5200.}$ &           LB-f1-651\\

\enddata
\end{deluxetable}
}
\normalsize

\end{document}